\documentclass[onecolumn,article,showpacs,nofootinbib,aps,prd,11pt]{revtex4}
\usepackage{graphicx,dcolumn,bm,amsfonts,amsmath,color,xcolor,amsthm}
\usepackage[colorlinks=true,citecolor=blue, linkcolor=blue,urlcolor=blue]{hyperref}
\usepackage{slashed}
\usepackage{ulem}
\usepackage{xcolor}

\newcommand{\blueSection}[1]{Section~\textcolor{blue}{#1}}
\allowdisplaybreaks
\begin{document}
\title{Addressing the lightest $S$-wave strange $K_0^*(700)/\kappa$ resonance in four-body semileptonic $\bar{B}_s^0 \to  K^0\pi^+ \ell^-\bar{\nu}_\ell$ decays}
\author{Dong Huang$^*$}
\address{Department of Physics, Guizhou Minzu University, Guiyang 550025, P.R. China}
\author{Sheng-Bo Wu\footnote{Dong Huang and Sheng-Bo Wu contributed equally to this work.}}
\address{Department of Physics, Guizhou Minzu University, Guiyang 550025, P.R. China}
\author{Fang-Ping Peng}
\address{Department of Physics, Guizhou Minzu University, Guiyang 550025, P.R. China}
\author{Long Zeng}
\address{Department of Physics, Guizhou Minzu University, Guiyang 550025, P.R. China}
\author{Hai-Bing Fu}
\email{fuhb@cqu.edu.cn}
\address{Department of Physics, Guizhou Minzu University, Guiyang 550025, P.R. China}

\begin{abstract}
As the lightest strange scalar resonance, $K_0^*(700)$ (also called $\kappa$) has a large width and resides close to the $K\pi$ threshold, leading to a longstanding debate about its internal structure. Within the framework of the conventional quark-antiquark ($q\bar{q}$) picture, this paper attempts to research the behaviors of $K_0^*(700)$ resonance in the four-body final state decays. Firstly, we investigates the $K_0^*(700)$ resonance one twist-2 and two twist-3 light-cone distribution amplitudes (LCDAs), {\it i.e.} $\phi_{2;K_0^*(700)}(x, \mu)$ and $\phi_{3;K_0^*(700)}^{p;\sigma}(x, \mu)$ by constructing light-cone harmonic oscillator model. For the longitudinal distribution characteristics of the twist-2 LCDA, two typical parametrization schemes, {\it i.e.} $\varphi_{2;K_0^*(700)}^{\rm (S1)}(x)$ and $\varphi_{2;K_0^*(700)}^{\rm (S2)}(x)$, are adopted in this paper. Meanwhile, it has enriched theoretical predictions for the first ten-order LCDAs $\xi$-moments. Subsequently, we apply the QCD light-cone sum rules approach to compute the $\bar B_s^0 \to K_0^*(700)$ transition form factors (TFFs) $f_\pm^{\bar B_s^0 K_0^*(700)}(q^2)$ and $f_{\rm T}^{\bar B_s^0 K_0^*(700)}(q^2)$ by taking into account the perturbative $\mathcal{O}(\alpha_s)$ corrections to the twist-2 LCDAs terms. By using the simplified series expansion parameterization, these TFFs are extrapolated over the full physical region. Then, we calculate the differential decay widths and branching fractions for four-body semileptonic decay $\bar{B}_s^0 \to K^0\pi^+ \ell^-\bar{\nu}_\ell$ with $\ell= (e, \mu, \tau)$ via the $S$-wave $K_0^*(700)$ resonance, respectively. We expect that the obtained prediction results will provide valuable insights for future experiments and theoretical research.
\end{abstract}

\maketitle

\section{INTRODUCTION}
To understand the nature of light scalar state with masses below $1\,{\rm GeV}$ has become a central issue within the realm of non-perturbative quantum chromodynamics (QCD). The reason is that light scalar states play a significant role in the hadron mass generation through the spontaneous breaking of QCD chiral symmetry. Consequently, understanding the confinement necessitates revealing of the internal quark-gluon substructure of these states. The existence of light scalar states below $1\,{\rm GeV}$, including isoscalars $f_0(500)/\sigma$, $f_0(980)$, isodoublet $K_0^*(700)/\kappa$ and isovector $a_0(980)$ states have been firmly established. In order to have a clear look at the scalar states with its quantum numbers, decay modes, branching fraction, mass and widths, those values are present in Table~\ref{table:meson information} from PDG~\cite{ParticleDataGroup:2024cfk}. The $K_0^*(700)$ and $f_0(500)$ states are have large decay widths, which are about $\Gamma_{K_0^*(700)}=(463\pm27)\,{\rm MeV}$ and $\Gamma_{f_0(500)}=(100-800)\,{\rm MeV}$. Meanwhile, their poles lie deep in the complex energy plane, which lead to substantial overlap among neighboring resonances or non-resonance backgrounds~\cite{Pelaez:2021dak, Close:2002zu}. The widths of $f_0(980)$ and $a_0(980)$ states are somewhat relatively small, which are approximately equal to $\Gamma_{f_0(980)} = (10-100)\,{\rm MeV}$ and $\Gamma_{a_0(980)} = (50-100)\,{\rm MeV}$. Their mass are close to $K\bar K$ threshold at around $990\,\rm{MeV}$. Due to the existence opening of multiple decay channels, the interference effect between these channels will producing cusps in the shapes of nearby resonances~\cite{Bugg:2004xu}. The occurrence of the above phenomena makes it difficult to clearly identify, separate and confirm the existence of scalar resonance states as an independent particle state and their physical parameters experimentally.
\begin{table}[htb]
\footnotesize
\begin{center}
\renewcommand{\arraystretch}{1.2}
\setlength{\tabcolsep}{16pt}
\caption{The properties of the scalar quark-antiquark scalar states below $1\,{\rm GeV}$ and their decay modes.}
\label{table:meson information}
\begin{tabular}{c l l c c c}
\hline
$\rm {Mesons}$&$J^{PC}$ & Decay Modes &$\mathcal{B}(\%)$&$\rm{Mass}$&$\rm{Width}$\\
\hline
$S$  &$$  &$M_{1}M_{2}$  &$S\to M_{1}M_{2}$  &${\rm MeV}$  &${\rm MeV}$
\\\hline
$K_0^*(700)$ &$0^+$     &$K\pi$                                   &$\sim100$   &$838\pm11$   &$463\pm27$ \\
$f_0(500)$   &$0^{++}$  &$\pi\pi/\gamma\gamma$                    &$\rm seen$  &$400-800$    &$100-800$  \\
$f_0(980)$   &$0^{++}$  &$\pi\pi/K\bar{K}/\gamma\gamma$           &$\rm seen$  &$990\pm20$   &$10-100$   \\
$a_0(980)$   &$0^{++}$  &$\eta\pi/K\bar{K}/\eta'\pi/\gamma\gamma$ &$\rm seen$  &$980\pm20$   &$50-100$   \\
\hline
\end{tabular}
\end{center}
\end{table}

Although the QCD theory has achieved remarkable success in describing strong interactions, the fundamental nature of light scalar mesons remains controversial~\cite{Godfrey:1998pd, Jaffe:2004ph, Pennington:2007eg}. In particular, various interpretations have been proposed, ranging from conventional quark-antiquark $(q\bar{q})$ states, tetraquarks $(qq\bar{q}\bar{q})$ states, meson-meson bound states or even supplemented with a scalar glueball. It is worth noting that scalar mesons and QCD vacuum have the same spin parity quantum numbers $(J^{PC}=0^{++})$. This means that scalar mesons are highly prone to mixing with the vacuum structure, thereby introducing strong uncertainly in the classification of their structures. Furthermore, the coupling constants of QCD in the low-energy region are often very large, approximately several $\rm GeV$, which makes the perturbation method unreliable. Therefore, understanding the internal structure of scalar mesons not only serves as the cornerstone for testing non-perturbative QCD, but also drives the research on non-perturbative QCD effects, multi-quark dynamics and hadron molecular states.

Due to long-standing ambiguity over the internal structure of scalar mesons, experimental and theoretical research are dedicated to conducting in-depth investigations into nature of the mesons through highly precise multi-body semileptonic decay processes. In particular, four-body decay channels involving scalar mesons offer a powerful avenue for investigation. The four-body weak decay processes offer a uniquely clean and powerful laboratory for investigating the internal structure of scalar resonance. This ``clean'' decay system involves only one hadron flow and one lepton flow, and the strong interaction effect is mainly encapsulated within the form factor, thereby providing an ideal environment for the precise extraction of fundamental parameters. It is worth noting that the transition form factors (TFFs) and branching fraction of the weak decay process are highly sensitive to the internal structure of scalar resonance, making the study of their dynamics a crucial tool for revealing the nature of scalar mesons. This endeavor bridges the gap between precision measurements of weak decays and strong interaction theories~\cite{Black:1998wt, Soni:2020sgn}.
\begin{table}[htb]
\footnotesize
\begin{center}
\renewcommand{\arraystretch}{1.2}
\setlength{\tabcolsep}{25pt}
\caption{Experimental status of light scalar resonances $f_0(500)$, $f_0(980)$ and $a_0(980)$ in four-body semileptonic decay processes.}
\label{decay channel}
\begin{tabular}{l l}
\hline
Decay Modes &Experimental  \\
\hline
$D^+\to f_0(500)(\to \pi^+\pi^-)e^+\nu_e$      &BESIII~\cite{BESIII:2018qmf, BESIII:2024lnh}    \\
$D^+\to f_0(500)(\to \pi^+\pi^-)\mu^+\nu_\mu$  &BESIII~\cite{BESIII:2024lnh}    \\
$D_s^+\to f_0(500)(\to \pi^+\pi^-)e^+\nu_e$    &BESIII~\cite{BESIII:2023wgr}    \\
$D_s^+\to f_0(500)(\to \pi^0\pi^0)e^+\nu_e$    &BESIII~\cite{BESIII:2021drk, BESIII:2023wgr}    \\
\hline
$D^+\to f_0(980)(\to \pi^+\pi^-)e^+\nu_e$    &BESIII~\cite{BESIII:2018qmf}  \\
$D_s^+\to f_0(980)(\to \pi^+\pi^-)e^+\nu_e$  &BESIII~\cite{BESIII:2023wgr}, CLEO~\cite{CLEO:2009dyb, CLEO:2009ugx} \\
$D_s^+\to f_0(980)(\to \pi^0\pi^0)e^+\nu_e$  &BESIII~\cite{BESIII:2021drk} \\
\hline
$D^0\to a_0(980)(\to \eta\pi^-)e^+\nu_e$     &BESIII~\cite{BESIII:2024zvp}\\
$D_s^+\to a_0(980)(\to \eta\pi^0)e^+\nu_e$   &BESIII~\cite{BESIII:2021tfk}\\
\hline
\end{tabular}
\end{center}
\end{table}
Furthermore, the invariant mass spectrum of the hadronic system can clearly resolve intermediate resonant structures while combined analyses of hadronic form factors and dynamical models enable discrimination among competing structural hypotheses~\cite{LHCb:2016nsl}. Such studies thus provide crucial experimental constraints for understanding the non-perturbative dynamics underlying the scalar meson spectrum. In recent years, systematic studies were conducted on the semileptonic decay processes of light scalar mesons, such as $f_0(500)$, $f_0(980)$ and $a_0(980)$, in experiments. As shown in Table~\ref{decay channel}, collaboration groups such as BESIII and CLEO have precisely measured multiple four-body final state precesses. The obtained results not only confirmed the existence of scalar resonances, but also provided important evidence for understanding the internal structure of scalar state resonances and the dynamics of hadrons. In contrast, no dedicated experimental study has yet been reported for the scalar $K_0^*(700)/\kappa$ resonance in similar four-body weak decay process.

Here, it is worth to focus on the light strange $K_0^*(700)$ state. Due to the mass of $K_0^*(700)$ state is less than $1\,\rm GeV$, there has always been considerable controversy regarding its internal quark structure. The various competing theoretical interpretations have been proposed to explain its nature. The $K_0^*(700)$ state has been analyzed under the quarkonium $(q\bar{q})$ state configuration in Refs.~\cite{Cheng:2005nb, Wang:2014vra, Li:2008tk, Issadykov:2015iba, Wang:2014upa, Sun:2010nv}. Other theoretical interpretations suggest that these states possess composite structures, such as compact tetraquarks $(q\bar{q}q\bar{q})$~\cite{Jaffe:1976ig, Jaffe:1976ih, Weinstein:1982gc, Maiani:2004uc, tHooft:2008rus, Kim:2017yvd, Ebert:2008id, Eichmann:2015cra, Francesco:0605191, Agaev:2018fvz}. Consequently, a comprehensive understanding of the $K_0^*(700)$ state fundamental properties demands enhanced experimental precision, advanced non-perturbative QCD techniques, and continued theoretical investigations.

In 1977, Jaffe first predicted the existence of the $K_0^*(700)$ state~\cite{Jaffe:1976ig}. Based on the MIT bag model, he identified a $0^+$ nonet consisting of an octet and a singlet within tetraquark system, corresponding to the known light scalar mesons. Among them, the $K_0^*(700)$ state resonance appears as the member with strangeness $S = 1$ and is extremely broad due to its decay to $K\pi$ via the ``fall-apart'' mechanism. This work was the first to suggest that light scalar mesons may not be conventional $q\bar{q}$ $P$-wave states and established that the existence of the $K_0^*(700)$ state is crucial for the interpretation of the entire nonet, thereby guiding subsequent experimental discoveries. Subsequently, S. Descotes-Genon et al. rigorously confirmed the existence of the $K_0^*(700)$ resonance by demonstrating the presence of a pole in the $\pi K \to \pi K$ amplitude on the second Riemann sheet. By analyzing the domain of validity of Roy-Steiner representations in the complex energy plane and selecting a representation that remains valid over a sufficiently wide region in the imaginary direction, they computed the $l = 0$ partial wave based solely on experimental data, without relying on model assumptions or additional approximations. The mass and width of the $K_0^*(700)$ resonance were determined as $m_{K_0^*(700)} = (658 \pm 13) \, \text{MeV}$ and $\Gamma_{K_0^*(700)} = (557 \pm 24) \, \text{MeV}$. This result provides a rigorous dispersive theory foundation for establishing the $K_0^*(700)$ as a genuine scalar resonance~\cite{Descotes-Genon:2006sdr}.

The above results originated from theoretical predictions until 2002, when the Fermilab E791 Collaboration~\cite{E791:2002xlc} provided clear experimental evidence. Based on an analysis of the Dalitz plot using 15, 090 events of the decay $D^+ \to K^- \pi^+ \pi^+ $, the collaboration found that a model containing only known $ K\pi$ resonances could not adequately fit the data, whereas the inclusion of an additional scalar resonance led to a significant improvement in the fit quality. The mass and width of this resonance were determined to be $m_{K_0^*(700)}=(797 \pm 19 \pm 43)\,{\rm MeV}$ and $\Gamma_{K_0^*(700)}=(410 \pm 43 \pm 87)\,{\rm MeV}$, respectively, while the parameters of the $K_0^*(1430)$-resonance were also redetermined as $m_{K_0^*(1430)}=(1459 \pm 7 \pm 5)\,{\rm MeV} $ and $\Gamma_{K_0^*(1430)}=(175 \pm 12 \pm 12)\,{\rm MeV} $. These results provide important experimental support for the study of scalar state spectroscopy. Subsequently, the FOCUS Collaboration~\cite{FOCUS:2007mcb} performed a Dalitz plot analysis based on 53, 653 events of the decay $D^+ \to K^- \pi^+ \pi^+ $, achieving the highest statistical precision to date for this channel and providing important experimental support for the theoretical confirmation described above. The analysis reveals a significant contribution from the $K_0^*(700)$ resonance in this decay process, with a Breit-Wigner fit fraction of $22.4\%$ and mass and width determined to be $m_{K_0^*(700)}=856 \pm 17 \pm 5 \pm 12 \, {\rm MeV}$ and $\Gamma_{K_0^*(700)}=464 \pm 28 \pm 6 \pm 21 \, {\rm MeV}$, respectively. These results are consistent with earlier measurements from experiments such as E791, further consolidating the key role of the $K_0^*(700)$ as a scalar resonance in the $K\pi$ system. In 2008, the CLEO Collaboration~\cite{CLEO:2008jus} performed a more refined study of the $K_0^*(700)$ resonance using a higher-statistics sample of $D^+ \to K^- \pi^+ \pi^+$ decays. By employing a quasi-model-independent approach, directly extracted the magnitude and phase of the $K\pi$ $S$-wave from threshold up to the kinematic limit, revealing that the $K_0^*(700)$ resonance plays a dominant role in this partial wave. This analysis not only confirmed the findings of earlier experiments such as FOCUS but also provided more precise experimental constraints for understanding the nature of light scalar mesons. Based on the experimental and theoretical analyses above, the dominant role of the $K_0^*(700)$ resonance in the $K\pi$ $S$-wave has been firmly established.

The branching fraction and TFFs rely on the assumption about the internal structure of the $K_0^*(700)$ resonance in the four-body $\bar{B}_s^0 \to K_0^*(700)(\to K^0 \pi^+) \ell^-\bar{\nu}_\ell$ decay. Different theoretical models may yield substantially different predictions for the TFFs of the $K_0^*(700)$ resonance, while calculations of the branching fractions for the $\bar{B}_s^0 \to K_0^*(700)(\to K^0\pi^+)\ell^-\bar{\nu}_\ell$ are highly sensitive to the underlying model assumptions. Therefore, an accurate calculation of these TFFs is crucial for a deeper understanding of the internal structure and dynamical mechanism of the $K_0^*(700)$ resonance. To date, under the $q\bar{q}$ state configuration, various theoretical groups have calculated the TFFs and branching fractions using different approaches, including the perturbative QCD (PQCD)~\cite{Li:2008tk}, the QCD light-cone sum rule (LCSR)~\cite{Wang:2014upa, Wang:2014vra, Sun:2010nv}, the covariant quark model (CQM)~\cite{Issadykov:2015iba}. To explore the differences among various theoretical predictions, we employ the LCSR to conduct theoretical predictions for the TFFs of the $\bar{B}_s^0 \to K_0^*(700)$ decays. The LCSR approach is an powerful tool for extracting and calculating non-perturbative parameters of various hadronic states in low-energy hadron physics. The amplitude of the $\bar{B}_s^0 \to K_0^*(700)$ transition can be factorized into a perturbatively calculable short-distance part and a non-perturbative long-distance part, which is parameterized by light-cone distribution amplitudes (LCDAs). Within this framework, the calculation involves performing an operator product expansion (OPE) near the light-cone ($x^2\approx 0$). The resulting non-perturbative hadronic matrix elements are then expressed in terms of LCDAs of different twists. The twist-2 LCDA $\phi_{2;K_0^*(700)}(u, \mu)$ of the $K_0^*(700)$ state has been systematically investigated using QCDSR~\cite{Cheng:2005nb}. By precisely calculating the first and third $\xi$-moments and the corresponding Gegenbauer moments, the study applied these results to the truncated form (TF) model of the Gegenbauer polynomial expansion. This approach not only determined the expansion coefficients but also successfully predicted the evolution behavior and shape characteristics of the $K_0^*(700)$ state LCDAs at different energy scales. Simultaneously, the twist-3 LCDAs were also investigated using their asymptotic forms, namely $\phi_{3;K_0^*(700)}^p(u, \mu)=\bar{f}_{K_0^*(700)}$ and $\phi_{3;K_0^*(700)}^\sigma(u, \mu)=\bar{f}_{K_0^*(700)}6u(1-u)$. However, our analysis of phenomenological models for the $\pi$ meson twist-2 LCDA reveals that the simple TF model has limitations in describing the complete dynamical behaviors of LCDAs~\cite{Zhong:2022lmn}. In the paper, we will adopt the systematic research framework previously developed for the $\pi$ meson twist-2 LCDA~\cite{Zhong:2021epq}, which has been successfully applied to precise calculations of twist-2 LCDA for the scalar $K_0^*(1430)$~\cite{Yang:2024jlz, Yang:2024ang, Huang:2022xny}, $a_0(980)$~\cite{Wu:2022qqx, Wu:2022gpy} and $a_0(1450)$ state~\cite{Song:2025mfm}. Specifically, we will construct LCHO model for $\phi_{2;K_0^*(700)}(u, \mu)$ and $\phi_{3;K_0^*(700)}^{p;\sigma}(u, \mu)$ based on the Brodsky-Huang-Lepage (BHL) prescription~\cite{BHL}, with model parameters determined through rigorous constraint conditions. This approach not only effectively overcomes the limitations of traditional truncated expansion methods but also provides reliable non-perturbative inputs for various high-energy processes involving mesons, thereby enhancing the precision of theoretical predictions.

This paper is organized as follows. In Sec.~\ref{II}, we first briefly introduce the decay width of the four-body semileptonic $\bar{B}_s^0 \to K_0^*(700)(\to K^0\pi^+)\ell^-\bar{\nu}_\ell$ decay. Then, we present the calculations of the TFFs for $B_s \to K_0^*(700)$ decays within the LCSR method and construct the twist-2,3 LCDAs $\phi_{2;K_0^*(700)}(x, \mu)$ and $\phi_{3;K_0^*(700)}^{p;\sigma}(x, \mu)$ within the LCHO model. In Sec.~\ref{III}, we present the relevant numerical results and compare them with other theoretical predictions. Section~\ref{IV} is reserved for a summary.

\section{THEORETICAL FRAMEWORK}\label{II}
\begin{figure}[htb]
\begin{center}
~\includegraphics[width=0.5\textwidth]{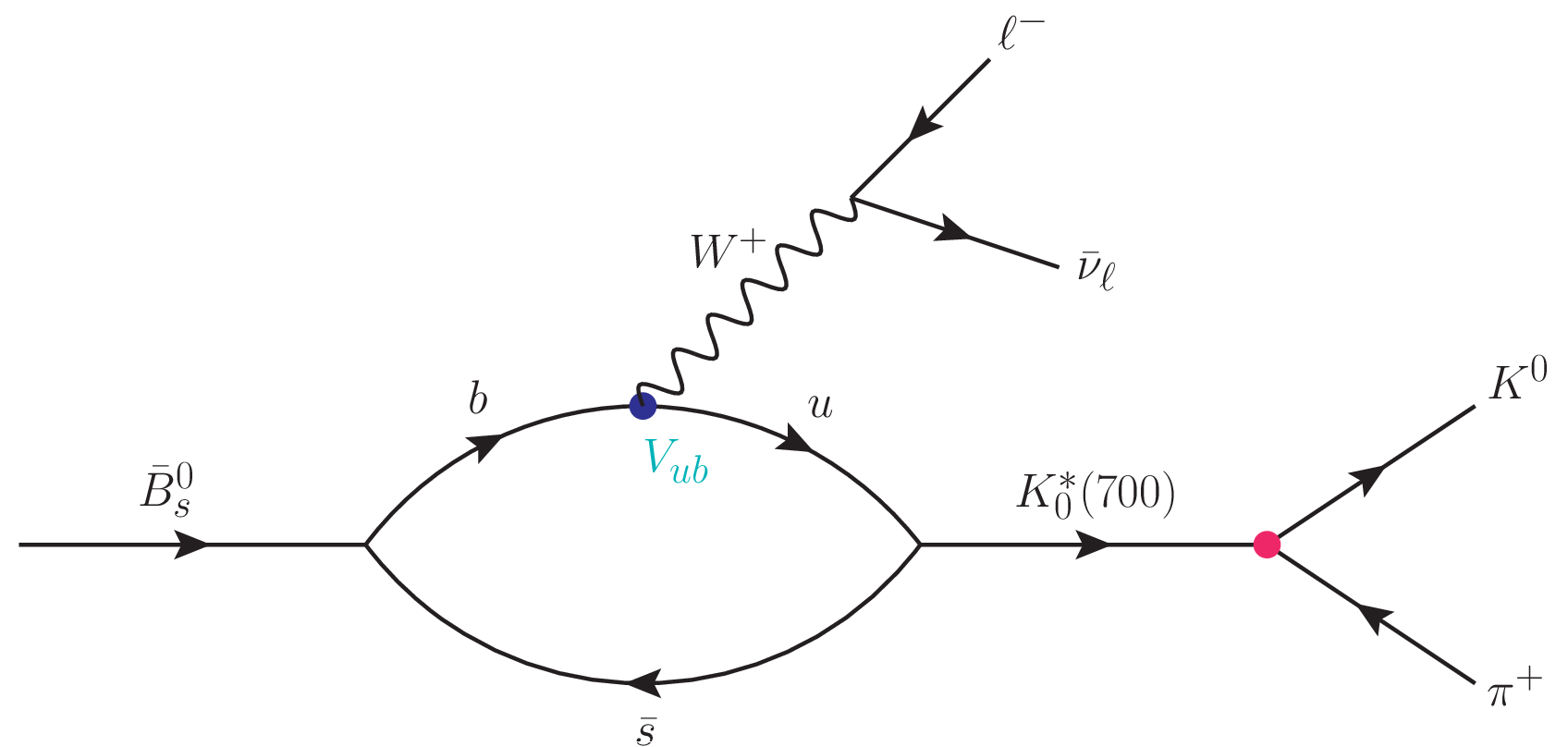}\\
\end{center}
\caption{The $\bar{B}_{s}^{0} \to K_0^*(700)(\to K^0\pi^+)\ell^-\bar{\nu}_\ell$ decay diagram, where the weak charged-current $b \to u W^+$ transition is dominated by the CKM matrix element $V_{ub}$, and the strange $K_0^*(700)$ resonance subsequently decays into the $K^0 \pi^+$ final state.}
\label{fig:FM}
\end{figure}
The decay diagram at the hadronic level for the process $\bar{B}_{s}^{0} \to K_0^*(700)(\to K^0\pi^+)\ell^-\bar{\nu}_\ell$ is shown in Fig.~\ref{fig:FM}.
The decay is dominated by the weak charged-current $b \to u W^+$ transition, with its coupling strength governed by the Cabibbo-Kobayashi-Maskawa (CKM) matrix element $V_{ub}$. After emitting an off-shell $W^+$-boson, the $u$-quark produced in the transition combines with the spectator $\bar{s}$ in the initial $\bar{B}_s^0$ meson to form the strange $K_0^*(700)$ resonance, which subsequently decays into the two-body final state $K^0\pi^+$ via the strong interaction. Meanwhile, the intermediate $W^+$-boson decays leptonically into a charged lepton $\ell^-$ and the associated antineutrino $\bar{\nu}_\ell$.

At the energy scale near $1\,{\rm GeV}$, the strange $K_0^*(700)$ resonance, due to its inherent instability, exhibits significant coupling with the $K\pi$ channel via the strong interactions. Given this feature, the intermediate resonance approximation can be adopted to describe the four-body semileptonic decay process. Under this approximation, the four-body decay process is equivalent to a three-body decay dominated by the intermediate $K_0^*(700)$ resonance. As a result, the decay amplitude can be factorized into two components, namely the three-body decay amplitude $\hat{A}$ and the hadronic decay amplitude of the intermediate resonance $K_0^*(700) \to K\pi$. The latter is described by the Breit-Wigner propagator together with the strong coupling vertex. Accordingly, the total decay amplitude for $\bar{B}_s^0 \to K_0^*(700)\ell^-\bar{\nu}_\ell\equiv K\pi\ell^-\bar{\nu}_\ell$ is expressed as~\cite{Wang:2016wpc}
\begin{align}
\mathcal{A}(\bar{B}_s^0 \to K\pi\ell^-\bar{\nu}_\ell) = \hat{A} \bigg(\frac{i}{D_{K_0^*(700)}}\times ig_{K_0^*(700)K\pi}\bigg),
\end{align}
where $g_{K_0^*(700) K\pi}$ represent the strong coupling constant of $K_0^*(700)\to K\pi$ vertex, and its particular isospin channel, such as $K^0\pi^+$ or $K^+\pi^0$, implicitly accounted for in the process-dependent normalization. The core component of the hadronic decay amplitude $\hat{A}$ is the weak transition matrix element for $\bar{B}_s^0 \to K_0^*(700)$ transition, whose Lorentz structure can be fully parameterized in terms of the TFFs~\cite{Huang:2022xny}:
\begin{align}
\langle K_0^*(700)(p)|\bar{u}\gamma_\mu\gamma_5b|\bar{B}_s^0(p+q) \rangle &= -2if_+^{\bar B_s^0 K_0^*(700)}(q^2)p_\mu  -  i[f_+^{\bar B_s^0 K_0^*(700)}(q^2) + f_-^{\bar B_s^0 K_0^*(700)}(q^2)]q_\mu,\label{eq:HME1}
\\
\langle K_0^*(700)(p)|\bar{u}\sigma_{\mu\nu}\gamma_5q^\nu b|\bar{B}_s^0(p+q)\rangle &= [2p_\mu q^2-2q_\mu (p\cdot q)]
\frac{-f_{\rm T}^{\bar B_s^0 K_0^*(700)}(q^2)}{m_{\bar{B}_s^0}+m_{K_0^*(700)}}.
\label{eq:HME2}
\end{align}
Here, the axial-vector current matrix element is described by $f_\pm^{\bar B_s^0 K_0^*(700)}(q^2)$, while the tensor current matrix element is characterized by $f_{\rm T}^{\bar B_s^0 K_0^*(700)}(q^2)$. When combining $\hat{A}$ with the strong coupling constant $g_{K_0^*(700)K\pi}$, the hadronic part of decay amplitude is fully determined by TFFs.
%[0512047]
The expression for the inverse propagator of $K_0^*(700)$ resonance, encompasses the self-energy correction, the width effect, and the resonance coupling information. It is typically expressed in the form~\cite{Achasov:2005hm,Achasov:2004uq}:
\begin{align}
D_{K_0^*(700)}(s) = m_{K_0^*(700)}^2\!-\!s+\!\sum_{K\pi}\Big[{\rm{Re}}\Pi_{K_0^*(700)}^{K\pi}(m_{K_0^*(700)}^2)-\Pi_{K_0^*(700)}^{K\pi}(s)\Big],
\end{align}
where the self-energy function $\sum_{K\pi}[\,{\rm{Re}}\Pi_{K_0^*(700)}^{K\pi}(m_{K_0^*(700)}^2)-\Pi_{K_0^*(700)}^{K\pi}(s)\,]$ takes into account the finite width corrections of $K_0^*(700)$ resonance, which are one-loop contributiond to the resonance self-energy from the two-particle intermediate $K\pi$ states subtracted at $s=m_{K_0^*(700)}^2$. The real part ${\rm Re}\Pi_{K_0^*(700)}^{K\pi}(m_{K_0^*(700)}^2)$ of self-energy function renormalizes the bare mass $m_{K_0^*(700)}$, shifting propagator pole away from the tree-level value. This pole close to a complex squared momentum, which defines the particle physical mass as an observable quantity in scattering experiments. After carrying out the Taylor expansion for the self-energy function, along with the constant term of renormalization mass, there is also a multiplicative constant from the derivative terms. These constants can be absorbed by normalizing the field operator via the field strength renormalization factor. The imaginary part ${\rm Im}\Pi_{K_0^*(700)}^{K\pi}(s)$ gives rise to a nonzero mass-dependent decay width $\Gamma_{K_0^*(700) K\pi}(s)$, reflecting the coupling strength of $K_0^*(700)$ resonance to the $K\pi$ continuum. The explicit form is as follows:
\begin{align}
{\rm Im}\:\Pi_{K_0^*(700)}^{K\pi}(s)=\sqrt{s}\;\Gamma_{K_0^*(700) K\pi}(s)=\frac{1}{16\pi}|g_{K_0^*(700) K\pi}|^2\rho_{K\pi}(s).
\end{align}
For pseudoscalar $K\pi$ mesons and $\sqrt{s}\geq m_+$, the self-energy function $\Pi_{K_0^*(700)}^{K\pi}(s)$ is defined as the following expression:
\begin{align}
\Pi_{K_0^*(700)}^{K\pi}(s)&=\frac{g_{K_0^*(700) K\pi}^2}{16\pi}\Bigg[\frac{m_+m_-}{\pi s}\:\ln\:\frac{m_{\pi}}{m_{K}}+\rho_{K\pi}(s)
\Bigg(i\!+\!\frac{1}{\pi}\ln\frac{\sqrt{s-m_-^2}\!-\!\sqrt{s-m_+^2}}{\sqrt{s-m_-^2}\!+\!\sqrt{s-m_+^2}}\Bigg)\Bigg],
\end{align}
where the individual phase space factor is given by Lorentz-invariant two-body phase space $\rho_{K\pi}(s)=\sqrt{(1-m_+^2/s)(1-m_-^2/s)}$ with $m_\pm=m_K\pm m_\pi$. Consequently, the denominator $D_{K_0^*(700)}$ can be expressed as:
\begin{align}
D_{K_0^*(700)} = m_{K_0^*(700)}^2-s-i\sqrt{s}\,\Gamma_{K_0^*(700) K\pi}(s),
\end{align}
Only the axial-vector current part contribution in the process of the $\bar{B}_s^0$ meson transition into the $K_0^*(700)$ resonance. Therefore, the double differential decay width takes the form:
\begin{align}
&\frac{d^2\Gamma(\bar{B}_s^0 \to K_0^*(700)(\to K^0\pi^+)\ell^-\bar{\nu}_{\ell})}{ds\,dq^2}=\frac{G_F^2|V_{ub}|^2}{192\pi^3 m_{\bar{B}_s^0}^3}
\lambda^{1/2} (m_{\bar{B}_s^0}^2, m_{K_0^*(700)}^2, q^2) \bigg(1 - \frac{m_\ell^2} {q^2}\bigg)^2\bigg[\bigg(1 + \frac{m_\ell^2}{2 q^2} \bigg)
\nonumber\\
&\qquad\times \lambda(m_{\bar{B}_s^0}^2, m_{K_0^*(700)}^2, q^2)|f_+^{\bar B_s^0 K_0^*(700)}(q^2)|^2+\frac{3m_\ell^2}{2q^2}(m_{\bar{B}_s^0}^2 -m_{K_0^*(700)}^2)^2|f_0^{\bar B_s^0 K_0^*(700)}(q^2)|^2\bigg]\frac{P(s)}{\pi},
\label{eq:DDW}
\end{align}
where $f_0^{\bar B_s^0 K_0^*(700)}(q^2)=f_+^{\bar B_s^0 K_0^*(700)}(q^2)+q^2/(m_{\bar{B}_s^0}^2-m_{K_0^*(700)}^2)f_-^{\bar B_s^0 K_0^*(700)}(q^2)$, $G_F$ is the Fermi coupling constant. The k\'allen function defined as $\lambda(x, y, z)=x^2+y^2+z^2-2xy-2xz-2yz$, and $P(s)$ is based on the relativistic Flatt\'e formula due to the open $K\pi$ channel as follows:
\begin{align}
P(s) = \frac{g_{K_0^*(700) K^0\pi^+}\rho_{K^0\pi^+}(s)}{|m_{K_0^*(700)}^2-s-ig_{K_0^*(700)K^0\pi^+}\rho_{K^0\pi^+}(s)|^2}.
\end{align}
Here, $s=(p_K+p_\pi)^2$ is the invariant mass squared $M_{K^0\pi^+}^2$ of the two pseudoscalars $K^0$, $\pi^+$. The invariant mass of final-state $K\pi$ system must satisfy $M_{K^0\pi^+}\geq(m_{K^0}+m_{\pi^+})$, as this is the threshold for forming an on-shell two-particle final state. Furthermore, in $\bar{B}_s^0 \to K^0\pi^+\ell^-\bar{\nu}_\ell$ decays, the invariant mass squared of lepton pair satisfies $q^2 \geq 0$, which implies that the invariant mass of $K\pi$ system is bounded by $\sqrt{s} \leq m_{\bar{B}_s^0}$.
The total decay width is~\cite{Dosch:2002rh}
\begin{align}
\Gamma=\int^{m_{\bar{B}_s^0}^2}_{(m_K+m_\pi)^2}ds\int^{(m_{\bar{B}_s^0}-m_{K_0^*(700)})^2}_{m_\ell^2}dq^2\;
\frac{d^2\Gamma(s, q^2)}{ds dq^2}.
\end{align}

The relevant TFFs can be calculated in the LCSR framework. For this purpose, we introduce two-point correlation functions, defined as the matrix elements of the T-ordered product of conventional current operators sandwiched between the vacuum state and an on-shell $K_0^*(700)$ state. More specifically, the form of vacuum-to-$K_0^*(700)$ relevant function we have adopted is as follows:
\begin{align}
\Pi_\mu(p, q) &=  i  \int  d^4x e^{iq\cdot x} \langle K_0^*(700)(p)| T\{ \bar{u}(x)\gamma_\mu \gamma_5 b(x)\bar{b}(0) i\gamma_5 s(0) \} |0\rangle
\nonumber\\
& = F(q^2, (p+q)^2)p_\mu + \widetilde{F}(p^2, (p+q)^2)q_\mu,
\label{eq:correlator1}
\\
\widetilde{\Pi}_\mu(p, q) &= i  \int d^4x e^{iq\cdot x} \langle K_0^*(700)(p)| T\{ \bar{u}(x)\sigma_{\mu\nu} \gamma_5 q^\nu b(x)
\bar{b}(0) i\gamma_5 s(0) \} |0\rangle
\nonumber\\
& = F^T(p^2, (p+q)^2) [p_\mu q^2 - q_\mu (p\cdot q)].
\label{eq:correlator2}
\end{align}
Here, $(p+q)$ denotes the four-momentum of the $B_s$ meson, while $p$ and $q$ correspond to the four-momentum and momentum transfer of $K_0^*(700)$ resonance, respectively. The central theoretical task is to evaluate the correlation functions in Eqs.~\eqref{eq:correlator1} and~\eqref{eq:correlator2} within QCD. This problem is most conveniently addressed in the Euclidean region. Under the kinematic conditions $q^2\ll m_b^2$ and $(p+q)^2\ll m_b^2$ (where $m_b$ denotes the $b$-quark mass), both momentum transfer scales lie well below the $b$-quark production threshold. The $b$-quark propagator in these correlation functions exhibits significant virtuality, with its dominant contributions confined to the light-cone region $x^2 \approx 0$~\cite{Khodjamirian:2005ea, Khodjamirian:2006st}. This feature establishes the theoretical foundation for the light-cone expansion of the heavy-quark propagator and the parametrization of non-perturbative hadronic structure via meson LCDAs. Therefore, this condition serves as the essential requirement for the realization of the factorization of perturbative and non-perturbative dynamics through the LCSR. Under these conditions, the OPE framework can employ Wick's theorem to contracts the $b$-quark fields, thereby expanding the vacuum-to-meson matrix elements into a series of meson LCDAs, and the series includes terms with increasing twist levels. As a first order approximation, we adopt the light-cone expansion formula of the $b$-quark propagator given in Ref.~\cite{Duplancic:2008ix}, which is given by
\begin{align}
\langle0|b^i_\alpha(x)\bar{b}^j_\beta(0)|0\rangle &= -i\int\frac{d^4k}{(2\pi)^4}e^{-ik\cdot x}\bigg\{\delta^{ij}\frac{\not\!{k}+m_b}{m_b^2-k^2}
+g_s\int_0^1dvG^{\mu\nu a}(vx)\Big(\frac{\lambda^a}{2}\Big)^{ij}\Big[\frac{\not\!{k}+m_b}{2(m_b^2-k^2)^2}\sigma_{\mu\nu}
\nonumber\\
&+\frac{1}{m_b^2-k^2}vx_\mu\gamma_\nu\Big]\bigg\}_{\alpha\beta}.
\label{eq:BP}
\end{align}
In our calculation, only the free quark propagator is retained in  Eq.~\eqref{eq:BP}, thereby truncating Fock components of the $K_0^*(700)$ state with parton multiplicities larger than two. Naturally, the vacuum-to-$K_0^*(700)$ state matrix element can be expanded in terms of the $K_0^*(700)$ state LCDAs of increasing twist. That is,
\begin{align}
\langle K_0^*(700)(p)\bar{u}^i_\alpha(x)s^j_\beta(0)|0\rangle
&=\frac{\delta^{ij}}{12} \bar{f}_{K_0^*(700)} \int^1_0 du e^{iup\cdot x} \Big[\not\!{p}\phi_{2;K_0^*(700)}(u,\mu) + m_{K_0^*(700)}\phi^p_{3;K_0^*(700)}(u,\mu)
\nonumber\\
&-\frac{1}{6}m_{K_0^*(700)}\sigma_{\mu\nu}p^{\mu}x^{\nu}\phi^\sigma_{3;K_0^*(700)}(u,\mu)\Big]_{\beta\alpha}+\cdots.
\end{align}
Here, $\bar{f}_{K_0^*(700)}$ denote the decay constant, $\phi_{2;K_0^*(700)}(u,\mu)$ and $\phi^{p;\sigma}_{3;K_0^*(700)}(u,\mu)$ represent the two-particle twist-2 and twist-3 LCDAs, respectively. Where $u$ is the fraction of the light-cone momentum of $K_0^*(700)$ carried by the $u$-quark, satisfying the normalization condition. By applying the OPE, the invariant amplitudes $F$, $\widetilde{F}$ and $F^T$ can be written in the following form,
\begin{align}
F_{\rm QCD} (q^2, (p+q)^2) &= i\bar{f}_{K_0^*(700)} \int^1_0\frac{du}{m_b^2-(up+q)^2}\bigg\{m_b\phi_{2;K_0^*(700)}(u,\mu)\! -\! um_{K_0^*(700)}\phi^p_{3;K_0^*(700)}(u,\mu)
\nonumber\\
&-\frac{m_{K_0^*(700)}}{6}\bigg[2+\frac{m_b^2+q^2-u^2p^2}{m_b^2 -(up+q)^2}\bigg]\phi^\sigma_{3;K_0^*(700)}(u,\mu)\bigg\},
\\
\widetilde{F}_{\rm QCD}(q^2, (p+q)^2)&=i\bar{f}_{K_0^*(700)} \int^1_0 \frac{du}{m_b^2-(up+q)^2}
\bigg\{-m_{K_0^*(700)} \phi_{3;K_0^*}^p(u,\mu)-\frac{1}{6u}\phi^\sigma_{3;K_0^*(700)}(u,\mu)
\nonumber\\
&\times  m_{K_0^*(700)} \bigg[1-\frac{m_b^2-q^2+u^2p^2}{m_b^2-(up+q)^2}\bigg]\bigg\},
\\
F_{\rm QCD} ^{\rm T} (q^2, (p+q)^2)&=\bar{f}_{K_0^*(700)}\int_0^1\frac{du}{m_b^2-(up+q)^2}
\bigg[\phi_{2;K_0^*(700)}(u,\mu)-\frac{m_b }{3[m_b^2-(up +  q)^2]}m_{K_0^*(700)}
\nonumber\\
&\times \phi^{\sigma}_{3;K_0^*(700)}(u,\mu)\bigg].
\end{align}
In the time-like $q^2$ region, following the standard LCSR procedure, the correlation functions in Eqs.~\eqref{eq:correlator1} and~\eqref{eq:correlator2} are decomposed into their hadronic spectral representation. We insert a complete set of intermediate hadronic states that sharing the quantum numbers of the interpolating currents, explicitly separate the single-particle pole contribution corresponding to the ground state $B_s$ meson, and denote the contribution from higher excited states as $\sum_{\rm H}$. In the calculation, the continuum contribution above the threshold $s_0^{B_s}$ is conventionally parameterized as the spectral integral. Consequently, the hadronic representation of the correlation function is written as:
\begin{align}
\Pi_{\mu}^{\rm had}(p, q)&=\frac{\langle K_0^*(700)(p)|\bar{u}\gamma_\mu\gamma_5b|B_s\rangle\langle B_s|\bar{b}i\gamma_5s|0\rangle}{m_{B_s}^2-(p+q)^2}
+\sum_{\rm H}\frac{\langle K_0^*(700)(p)|\bar{u}\gamma_\mu\gamma_5b|B_s^{\rm H}\rangle\langle B_s^{\rm H}|\bar{b}i\gamma_5s|0\rangle}{m_{B_s^{H}}^2-(p+q)^2}
\nonumber\\
&= F[q^2, (p+q)^2]p_\mu+\widetilde{F}[p^2, (p+q)^2]q_\mu.
\label{eq:HR}
\end{align}
For the $B_s$ meson, its vacuum-to-meson matrix element reads $\langle B_s|\bar{b}i\gamma_5s|0\rangle = m_{B_s}^2f_{B_s}/(m_b+m_s)$, with $f_{B_s}$ denote the $B_s$ meson decay constant. Substituting the parameterized expressions from Eqs.~\eqref{eq:HME1} and \eqref{eq:HME2} into Eq.~\eqref{eq:HR}, we neglect contributions from higher excited states, continuum states and possible subtraction terms. Therefore, the invariant amplitude $F_{\rm had}$  and $\widetilde{F}_{\rm had}$ can be written as
\begin{align}
F_{\rm had}(p^2, (p+q)^2)&=\frac{-2im_{B_s}^2f_{B_s}f_+^{\bar B_s^0 K_0^*(700)}(q^2)}{(m_b+m_s)[m_{B_s}^2-(p+q)^2]}
+\int_{m_b^2}^{s_0^{B_s}}ds\frac{\rho_+^{\alpha_s}(s)}{s-(p+q)^2}+\int_{s_0^{B_s}}^{\infty}ds\frac{\rho_+(s)}{s-(p+q)^2},
\nonumber\\
\widetilde{F}_{\rm had}(p^2, (p+q)^2)&=\frac{-im_{B_s}^2f_{B_s}\Big[f_+^{\bar B_s^0 K_0^*(700)}(q^2)+f_-^{\bar B_s^0 K_0^*(700)}(q^2)\Big]}{(m_b+m_s)[m_{B_s}^2-(p+q)^2]}
+\int_{m_b^2}^{s_0^{B_s}}ds\frac{\rho_\pm^{\alpha_s}(s)}{s-(p+q)^2}
\nonumber\\
&+\int_{s_0^{B_s}}^{\infty}ds\frac{\rho_\pm(s)}{s-(p+q)^2},
\nonumber\\
F_{\rm had}^{\rm T}(p^2, (p+q)^2)&=\frac{-2m^2_{B_s}f_{B_s}f_{\rm T}^{\bar B_s^0 K_0^*(700)}(q^2)}{(m_b+m_s)(m_{B_s}+m_{K_0^*(700)})[m_{B_s}^2-(p+q)^2]}+\int_{m_b^2}^{s_0^{B_s}}ds
\frac{\rho_T^{\alpha_s}(s)}{s-(p+q)^2}
\nonumber\\
&+\int_{s_0^{B_s}}^{\infty}ds\frac{\rho_T(s)}{s-(p+q)^2}.
\end{align}
We take the quark-hadronic duality approximation~\cite{Shifman:1978bx, Shifman:1978by} below the continuum thresholds $s_0^{B_s}$ and perform the Borel transformation on the squared momentum $(p+q)^2$ of the $B_s$ meson. The resulting LCSRs for the TFFs can be expressed as
\begin{align}
f_+^{\bar B_s^0 K_0^*(700)}(q^2) &= \frac{(m_b+m_{s})\bar{f}_{K_0^*(700)} }{2m_{B_{s}}^2 f_{B_{s}}} \exp\bigg[\frac{m_{B_{s}}^2}{M^2}\bigg]
\int^{\widetilde{u}_0}_{u_0}  du \exp  \bigg[-\frac{ m_b^2-\bar{u} q^2+u\bar{u} m_{K_0^*(700)}^2}{uM^2}\bigg]
\bigg\{\!\!-\! m_b
\nonumber\\
&\times \frac{\phi_{2;K_0^*}(u,\mu)}{u} + m_{K_0^*(700)}\:\phi_{3;K_0^*}^p(u,\mu) + m_{K_0^*(700)} \bigg[\;\frac{2}{u} + \frac{4u m_b^2 \,m_{K_0^*(700)}^2}{(m_b^2-q^2+u^2 \, m_{K_0^*(700)}^2)^2}
\nonumber\\
&- \frac{m_b^2+q^2-u^2m_{K_0^*(700)}^2}{m_b^2-q^2+u^2m_{K_0^*(700)}^2}\: \frac{d}{du}\; \bigg] \frac{\phi_{3;K_0^*}^\sigma(u,\mu)}{6} \bigg\}
\! + \! \frac{f_{\bar{K}_0^*}}{2m_{B_s}^{2}f_{B_s}}\exp\bigg[\frac{m_{B_s}^{2}}{M^2}\bigg]
\int_{m_b^2}^{s_0^{B_s}}ds\rho_{+}^{\alpha_s}(s)
\nonumber\\
&\times \exp\bigg[-\frac{s}{M^2}\bigg], \label{eq:LCSRTFFs1}
\\
f_+^{\bar B_s^0 K_0^*(700)}(q^2) &+ f_-^{\bar B_s^0 K_0^*(700)}(q^2) = \frac{(m_b+m_{s})\bar{f}_{K_0^*(700)}}{m_{B_{s}}^2 f_{B_{s}}} m_{K_0^*(700)} \exp\bigg[\frac{m_{B_{s}}^2}{M^2}\bigg]
\int^{\widetilde{u}_0}_{u_0} du
\bigg[ \frac{\phi_{3;K_0^*}^p(u,\mu)}{u}
\nonumber\\
& + \frac{1}{6u} \frac{d}{du}\:\phi_{3;K_0^*}^\sigma(u,\mu) \bigg] \exp\bigg[-\frac{m_b^2-\bar{u}\,q^2+u\,\bar{u}\,m_{K_0^*(700)}^2}{uM^2}\bigg]
+\frac{\bar{f}_{K_0^*(700)}}{m_{B_s}^{2}f_{B_s}} \: \exp\bigg[\frac{m_{B_s}^{2}}{M^2}\bigg]
\nonumber\\
& \times \int_{m_b^2}^{s_0^{B_s}} ds \rho_{+-}^{\alpha_s}(s)\exp\bigg[ - \frac{s}{M^2}\bigg],
\label{eq:LCSRTFFs2}
\\
f_{\rm T}^{\bar B_s^0 K_0^*(700)}(q^2) &= \frac{(m_b+m_{s}) \bar{f}_{K_0^*(700)}}{m_{B_{s}}^2 f_{B_{s}}} \: (m_{B_{s}} + m_{K_0^*(700)})\: \exp\bigg[\frac{m_{B_{s}}^2}{M^2}\bigg]
\int^{\widetilde{u}_0}_{u_0}du \: \bigg\{-\frac{\phi_{2;K_0^*}(u,\mu)}{2u}
\nonumber\\
& + \frac{m_b}{m_b^2-q^2+u^2m_{K_0^*(700)}^2} m_{K_0^*(700)} \bigg[ \frac{2um_{K_0^*(700)}^2}{m_b^2-q^2+u^2m_{K_0^*(700)}^2} \!-\! \frac{d}{du} \bigg]
\frac{\phi_{3;K_0^*}^\sigma(u,\mu)}{6} \bigg\}
\nonumber\\
&\times \exp\bigg[-\frac{m_b^2-\bar{u}q^2+u\bar{u}m_{K_0^*(700)}^2}{uM^2} \bigg] +\frac{(m_{B_s}+m_{K_0^*(700)})\bar{f}_{K_0^*(700)}}{2m_{B_s}^{2}f_{B_s}} \exp\bigg[\frac{m_{B_s}^{2}}{M^2}\bigg]
\nonumber\\
&\times \int_{m_b^2}^{s_0^{B_s}} ds \rho_{T}^{\alpha_s}(s)\exp\bigg[-\frac{s}{M^2}\bigg], \label{eq:LCSRTFFs3}
\end{align}
with
\begin{align}
u_0 &= \Big[ \sqrt{(q^2 - s_0^{B_s} + m_{K_0^*(700)}^2)^2 + 4m_{K_0^*(700)}^2 (m_b^2 - q^2)}+ q^2 - s_0^{B_s} + m_{K_0^*(700)}^2 \Big]/(2m_{K_0^*(700)}^2),
\nonumber\\
\widetilde{u}_0 &= \Big[ \sqrt{(q^2 - m_b^2 + m_{K_0^*(700)}^2)^2 + 4m_{K_0^*(700)}^2 (m_b^2 - q^2)}+ q^2 - m_b^2 + m_{K_0^*(700)}^2 \Big]/(2m_{K_0^*(700)}^2).
\end{align}
The explicit expressions for $\rho_{+}^{\alpha_s}(s)$, $\rho_{+-}^{\alpha_s}(s)$ and $\rho_{T}^{\alpha_s}(s)$ can be found in the Ref.~\cite{Wang:2014vra}. The LCDAs $\phi_{2;K_0^*(700)}(x, \mu)$ and $\phi_{3;K_0^*(700)}^{p;\sigma}(x, \mu)$ describe the momentum fraction distributions of the partons in the $K_0^*(700)$ state corresponding to the lowest Fock component. These universal non-perturbative quantities serve as essential inputs for the calculation of TFFs in exclusive processes $\bar{B}_s^0 \to K_0^*(700)$. For the investigation of these LCDAs, phenomenological model approaches can be employed. Based on the BHL prescription~\cite{BHL}, we construct a LCHO model to accurately describe the momentum fraction distribution of $\phi_{2;K_0^*(700)}(x, \mu)$ and $\phi_{3;K_0^*(700)}^{p;\sigma}(x, \mu)$. The LCHO model rigorously satisfies the theoretical constraints imposed by QCD on LCDAs, accurately describes their asymptotic behavior at the endpoints, and effectively suppresses endpoint singularities commonly encountered in certain processes. Furthermore, the LCHO model provides a scale evolution framework for LCDAs that bridges non-perturbative hadronic structure and perturbative QCD calculations. The core of the BHL prescription is to establish a theoretical bridge connecting two reference frames with different advantages in hadron physics, namely the rest frame and the light-cone frame. while these two frames differ significantly in their mathematical descriptions, they describe the same physical meson state. Accordingly, the BHL prescription starts from the assumption that there exists a connection between the equal-times wave function (WF) in the rest frame and the light-cone WF. In the hadrons bound state, constituent quarks do not behave as free particles but exist in off-shell configurations. The BHL prescription requires that the off-shell propagators in the rest frame and the light-cone frame share the same invariant mass distribution. Subject to this constraint, approximate bound state solutions from the quark model in the rest frame can be mapped to the light-cone frame, allowing us to construct the LCHO model for the meson WF. Formally, the light-cone WF for the $K_0^*(700)$ state, which enter its LCDAs, is given by
\begin{align}
\Psi_{i;K_0^*(700)}(x, \mathbf{k}_\perp) = \chi_{i;K_0^*(700)}(x, \mathbf{k}_\perp) \Psi^R_{i;K_0^*(700)}(x, \mathbf{k}_\perp).
\label{eq:WF}
\end{align}
Here, $i=2,3$ corresponds to the twist-2 and twist-3 LCDAs, respectively. The transverse momentum is denoted by $\mathbf{k}_\perp$, and $\chi_{i;K_0^*}(x, \mathbf{k}_\perp)$ is the spin-space WF arising from the Wigner-Melosh rotation. The spatial WF $\Psi^R_{i;K_0^*}(x, \mathbf{k}_\perp)$ can be decomposed into two parts: the $u-$dependent component $\varphi_{i;K_0^*}(x)$, which governs the longitudinal distribution of the WF, and the $\mathbf{k}_\perp-$dependent component derived from the harmonic oscillator solution for the meson in the rest frame. The $K_0^*(700)$ state WF and their corresponding LCDAs are related via the following integral equation:
\begin{align}
\phi _{i;K_0^*(700)}(x, \mu) = \int_{|\mathbf{k}_\perp|^2 \le \mu^2} \frac{d^2\mathbf{k}_\perp}{16\pi^3} \Psi_{i;K_0^*(700)}(x, \mathbf{k}_\perp).
\label{eq:DAWF}
\end{align}
After integrating over the transverse momentum $\mathbf{k}_\perp$, the LCDAs can be expressed as
\begin{align}
\phi_{2;K_0^*(700)}(x, \mu) &= \frac{A_{2;K_0^*(700)} \beta_{2;K_0^*(700)} \widetilde{m}}{4\sqrt{2} \,\pi^{3/2}}
\sqrt{x\bar{x}} \exp\bigg[-\frac{\hat{m}_q^2 x + \hat{m}_s^2 \bar{x} - \widetilde{m}^2}{8\beta_{2;K_0^*(700)}^2 x\bar{x}} \bigg]
\bigg\{ {\rm Erf} \bigg( \sqrt{\frac{\widetilde{m}^2 + \mu^2}{8\beta_{2;K_0^*(700)}^2 x\bar{x}}} \bigg)
\nonumber\\
& - {\rm Erf} \bigg( \sqrt{\frac{\widetilde{m}^2}{8\beta_{2;K_0^*(700)}^2 x\bar{x}}} \bigg) \bigg\}  \varphi_{2;K_0^*(700)}^{\rm (S1, S2)}(x),
\nonumber\\
\phi_{3;K_0^*(700)}^{p}(x, \mu) &= \frac{A_{3;K_0^*(700)}^{p} (\beta_{3;K_0^*(700)}^{p})^2}{2\pi^{2}} \exp\bigg[ \!\! - \! \frac{\hat{m}_q^2\bar{x}+\hat{m}_s^2x}{8(\beta_{3;K_0^*(700)}^{p})^2x\bar{x}}\bigg]
\bigg\{ \! 1 \! - \! \exp\bigg[\!\! - \! \frac{\mu^2}{8x\bar{x}(\beta_{3;K_0^*(700)}^{p})^2}\bigg]\bigg\}
\nonumber\\
& \times \varphi_{3;K_0^*(700)}^{p}(x),
\nonumber\\
\phi_{3;K_0^*(700)}^{\sigma}(x, \mu) &= \frac{A_{3;K_0^*(700)}^{\sigma} (\beta_{3;K_0^*(700)}^{\sigma})^2}{2\pi^{2}} \exp \bigg[\!\! - \! \frac{\hat{m}_q^2\bar{x}+\hat{m}_s^2 x}{8(\beta_{3;K_0^*(700)}^{\sigma})^2 x\bar{x}}\bigg]
\bigg\{\! 1 \! - \! \exp\bigg[\!\! -\! \frac{\mu^2}{8x\bar{x}(\beta_{3;K_0^*(700)}^{\sigma})^2}\bigg]\bigg\}
\nonumber\\
& \times \varphi_{3;K_0^*(700)}^{\sigma}(x). \label{eq:DAsigma}
\end{align}
Here, the error function is defened as ${\rm Erf}(x)=2\int_0^x dt e^{-t^2}/\sqrt{\pi}$, and $\widetilde{m}=\hat{m}_{q}x+\hat{m}_{s}\bar{x}$ with $\bar{x}=1-x$. The parameters $\hat{m}_q$ and $\hat{m}_s$ are the constituent quark masses of $K_0^*(700)$ state. For the longitudinal distribution functions of the twist-2 LCDA, $i.e.,$ $\varphi_{2;K_0^*}(x)$, we consider two different schemes. In the first scheme ${\rm (S1)}$, we take $\varphi_{2;K_0^*(700)}^{\rm (S1)}(x) =  C_1^{3/2}(2x - 1)$. In the second scheme ${\rm (S2)}$, we use $\varphi_{2;K_0^*(700)}^{\rm (S2)}(x) = (x\bar{x})^{\alpha_{2;K_0^*(700)}} C_1^{3/2}(2x - 1)$. For the $\varphi_{3;K_0^*(700)}^p(x)$ and $\varphi_{3;K_0^*(700)}^\sigma(x)$, we adopt the leading term of Gegenbauer expansion, {\it i.e.}, $\varphi_{3;K_0^*(700)}^p(x) = 1+\hat{B}_{1, p}^{3;K_0^*(700)}(\mu)C_1^{1/2}(2x - 1)$ and $\varphi_{3;K_0^*(700)}^\sigma(x) = 6x\bar{x}[1+\hat{B}_{1, \sigma}^{3;K_0^*(700)}(\mu)C_1^{3/2}(2x - 1)]$, where $C_1^{1/2;3/2}(2x-1)$ is the Gegenbauer polynomial. The LCHO model parameters are determined by four constraints, namely the average transverse momentum squared $\langle {\rm \mathbf{k}_{\perp}^2\rangle}$, the Gegenbauer moments $a_n^{2;K_0^*(700)}(\mu)$, the normalization condition satisfied by the twist-3 LCDAs $\phi_{3;K_0^*(700)}^{p;\sigma}(x, \mu)$, and the probability of finding the lowest Fock state $|u\bar{s}\rangle$ in the $K_0^*(700)$ expansion.

\section{NUMERICAL ANALYSIS}\label{III}
\subsection{The LCDAs $\phi_{2;K_0^*(700)}(x, \mu)$ and $\phi_{3;K_0^*(700)}^{p;\sigma}(x, \mu)$ behaviors}
\begin{figure}[htb]
\begin{center}
~\includegraphics[width=0.48\textwidth]{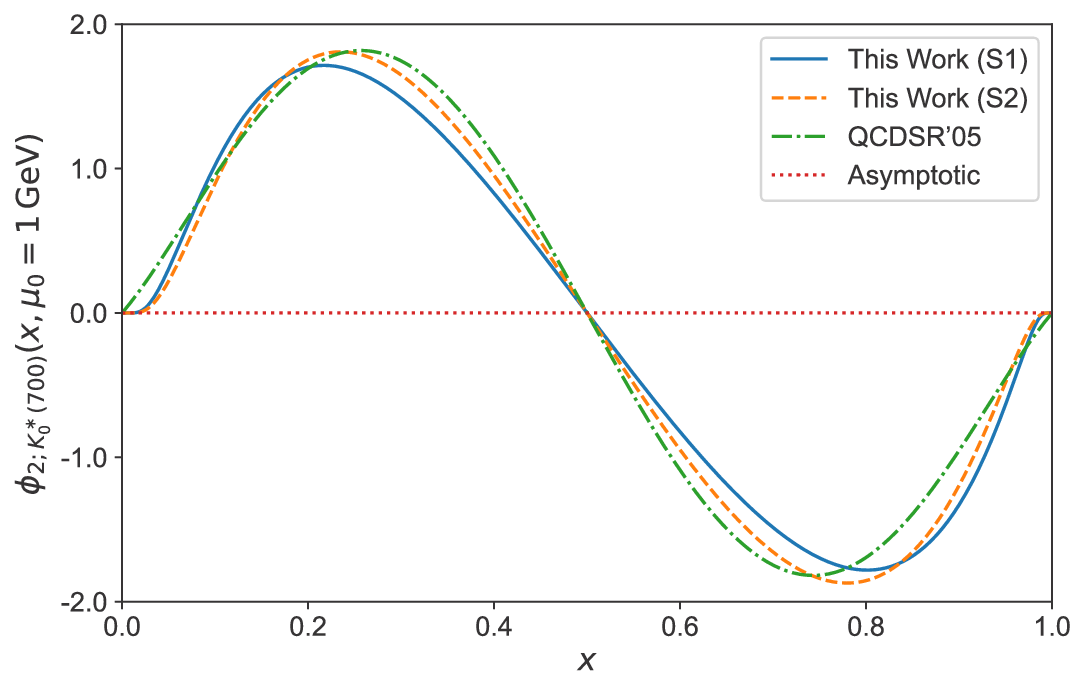}
~\includegraphics[width=0.48\textwidth]{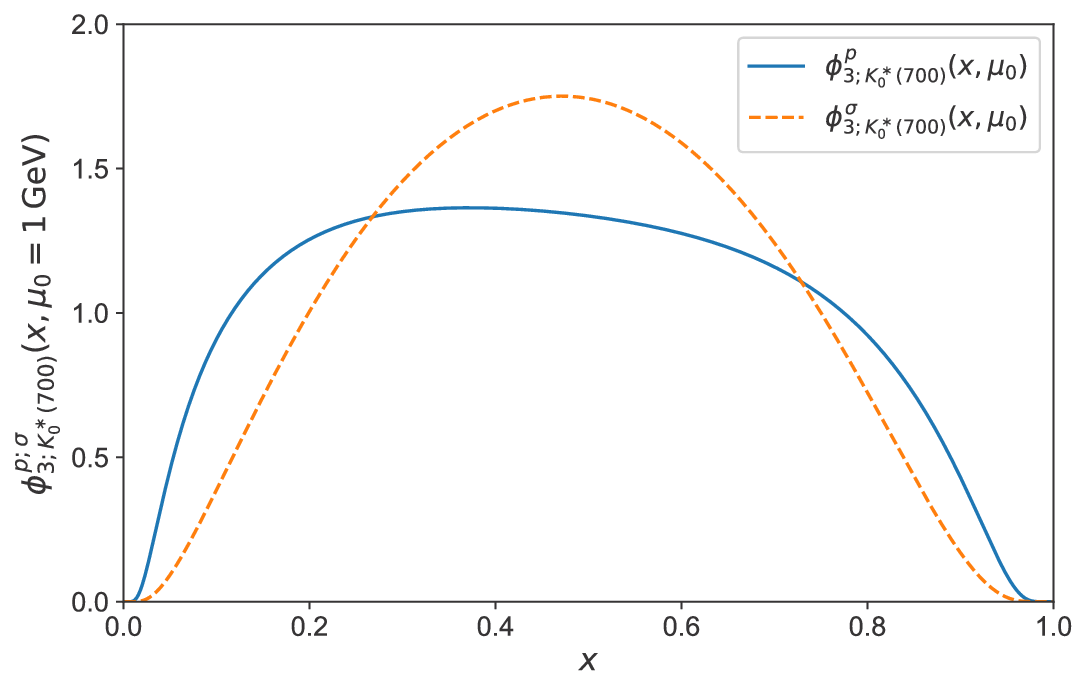}
\end{center}
\caption{Predictions for the behavior of the twist-2 LCDA $\phi_{2;K_0^*(700)}(x, \mu_0=1\,{\rm GeV})$ and the twist-3 LCDAs $\phi_{3;K_0^*(700)}^{p;\sigma}(x, \mu_0=1\,{\rm GeV})$ as function of the momentum fraction. For comparison, the twist-2 LCDA from the conventional QCDSR'05~\cite{Cheng:2005nb}, computed within the TF model framework, is also displayed.}
\label{fig:phi}
\end{figure}

According to the discussion in~\blueSection{II}, LCDAs $\phi_{2;K_0^*(700)}(x, \mu)$ and $\phi_{3;K_0^*(700)}^{p;\sigma}(x, \mu)$ constitute the crucial non-perturbative input parameters for the four-body semileptonic $\bar{B}_s^0 \to K_0^*(700)(\to K\pi)\ell^-\bar{\nu}_\ell$ decays. It is therefore essential to determine the LCDAs behaviors. In our calculations, we adopt the constituent quark masses $\hat{m}_s = 450\,{\rm MeV}$ and $\hat{m}_q = 300\,{\rm MeV}$~\cite{zhong:1109.3127}. Based on the LCHO model we constructed and taking into account the relevant theoretical constraints, we present the following parametrization considerations. For the ${\rm (S1)}$ scheme, we take $\langle {\rm \mathbf{k}_{\perp}^2\rangle}_{K_0^*(700)}^{1/2}=0.35\,{\rm GeV^2}$~\cite{Guo:1991eb} and $a_1(\mu_0)=-0.92 \pm 0.11$~\cite{Cheng:2005nb}. For the ${\rm (S2)}$ scheme, we additionally impose the condition $a_3(\mu_0)=0.15\pm0.09$. The LCHO model parameters of two twist-2 LCDA are defined at the initial scale $\mu_0=1\,\rm GeV$,
\begin{align}
A_{2;K_0^*(700)}^{\rm (S1)}&=-967.284, ~\beta_{2;K_0^*(700)}^{\rm (S1)}=0.558;\nonumber\\
A_{2;K_0^*(700)}^{\rm (S2)}&=-1600.14, ~\beta_{2;K_0^*(700)}^{\rm (S2)}=0.543, ~\alpha_{2;K_0^*(700)}^{\rm (S2)}=0.226.
\label{eq:T2}
\end{align}
For the twist-3 LCDAs $\phi_{3;K_0^*(700)}^{p;\sigma}(x, \mu_0)$, we set the squared transverse momentum $\langle {\rm \mathbf{k}_{\perp}^2\rangle}_{K_0^*(700)}^{1/2}=0.35\,{\rm GeV^2}$, together with the probability of finding the lowest Fock state $|u\bar{s}\rangle$ in the $K_0^*(700)$ state expansion. In this work, we take this probability to be $P_{K_0^*(700)}\approx 0.6$, inferred primarily from the prediction for the $K$-meson $P_K \approx 0.52$~\cite{Guo:1991eb}. Meanwhile, the normalization condition is employed to constrain the remaining parameters. The resulting model parameters for $\phi_{3;K_0^*(700)}^{p;\sigma}(x, \mu_0)$ are presented as follows,
\begin{align}
A_{3;K_0^*(700)}^{p} &= 268.739, ~\hat{B}_{1, p}^{3;K_0^*(700)} = 0.016, ~\beta_{3;K_0^*(700)}^p = 0.402;\nonumber\\
A_{3;K_0^*(700)}^{\sigma} &= 282.400, ~\hat{B}_{1, \sigma}^{3;K_0^*(700)} = 0.013, ~\beta_{3;K_0^*(700)}^{\sigma} = 0.375.
\label{eq:T3}
\end{align}
The LCDAs $\phi_{2;K_0^*(700)}^{{\rm (S1, S2)}}(x, \mu_0)$ and $\phi_{3;K_0^*(700)}^{p;\sigma}(x, \mu_0)$ versus the momentum fraction are shown in Fig.~\ref{fig:phi}. For comparison, the corresponding results from the conventional QCDSR approach~\cite{Cheng:2005nb} are also shown.
In Ref.~\cite{Cheng:2005nb}, the TF model for the twist-2 LCDA of $K_0^*(700)$ state is employed. Our results presented indicate that the twist-2 LCDA $\phi_{2;K_0^*(700)}^{{\rm (S1)}}(x, \mu_0)$ takes the maximum value at $x=0.217$ $(\phi_{2;K_0^*(700)}^{{\rm (S1)}}(x=0.217, \mu_0)\simeq 1.715)$ and reaches it minimum value at $x=0.801$ $(\phi_{2;K_0^*(700)}^{{\rm (S1)}}(x=0.801, \mu_0)\simeq -1.781)$. For scheme (S2), the corresponding extrema occur at $\phi_{2;K_0^*(700)}^{{\rm (S2)}}(x=0.235, \mu_0)\simeq 1.809$ and $\phi_{2;K_0^*(700)}^{{\rm (S2)}}(x=0.780, \mu_0)\simeq -1.871$, while both schemes cross zero point at $x = 0.5$. From Fig.~\ref{fig:phi}, one can also find that, our predicted behaviors for the twist-2 LCDAs are consistent with the theoretical result from QCDSR'05, both of which exhibiting antisymmetric characteristics. Using the results of Eq.~\eqref{eq:T3}, the behavior of twist-3 LCDAs can be obtained. To date, most previous studies of twist-3 LCDAs have employed the asymptotic forms, such as $\phi_{3;K_0^*(700)}^{p}(x)=\bar{f}_{K_0^*(700)}$ and $\phi_{3;K_0^*(700)}^{\sigma}(x)=\bar{f}_{K_0^*(700)}6x\bar{x}$~\cite{Cheng:2005nb, Wang:2014upa, Wang:2014vra, Sun:2010nv}. Motivated by this, we present the LCHO model for twist-3 LCDAs. Our results provide essential input parameters for exclusive processes involving the $K_0^*(700)$ state.

Using the LCDAs $\phi_{2;K_0^*(700)}(x, \mu_0)$ and $\phi_{3;K_0^*(700)}^{p, \sigma}(x, \mu_0)$, the $n$-th $\xi$-moment is difened as
\begin{align}
\langle\xi^n\rangle_{i;K_0^*(700)}(\mu_0)=\int_0^1dx\,(2x-1)^n\phi_{i;K_0^*(700)}(x, \mu_0).\label{eq:xi}
\end{align}
The first ten-order $\xi$-moments $\langle \xi^{n} \rangle_{2;K_0^*(700)}^{\rm (S1,S2)}(\mu_0)$ and $\langle \xi^{n} \rangle_{3;K_0^*(700)}^{p, \sigma}(\mu_0)$ at the scale $\mu_0=1\,{\rm GeV}$ are calculated and listed in Table~\ref{table:xin}.
\begin{table}[htb]
\footnotesize
\begin{center}
\renewcommand{\arraystretch}{1.2}
\setlength{\tabcolsep}{18pt}
\caption{Predictions for the first ten-order $\xi$-moments $\langle \xi^{n} \rangle_{2;K_0^*(700)}^{\rm (S1,S2)}(\mu_0)$ and $\langle \xi^{n} \rangle_{3;K_0^*(700)}^{p, \sigma}(\mu_0)$ with $n=(1, 2, \cdots, 10)$ of $K_0^*(700)$ state at the scale $\mu_0=1~{\rm GeV}$. Where all parameters are taken at their central values.}
\label{table:xin}
\begin{tabular}{l c c c c c}
\hline
$n$ &$\langle\xi^n\rangle_{2;K_0^*(700)}^{\rm (S1)}(\mu_0)$ &$\langle\xi^n\rangle_{2;K_0^*(700)}^{\rm (S2)}(\mu_0)$ &$\langle\xi^n\rangle_{3;K_0^*(700)}^p(\mu_0)$ &$\langle\xi^n\rangle_{3;K_0^*(700)}^\sigma(\mu_0)$ \\
\hline
$1$      &$-0.552$ &$-0.552$ &$-0.073$ &$-0.044$ \\
$3$      &$-0.220$ &$-0.208$ &$-0.038$ &$-0.018$ \\
$5$      &$-0.113$ &$-0.102$ &$-0.023$ &$-0.009$ \\
$7$      &$-0.066$ &$-0.058$ &$-0.016$ &$-0.005$ \\
$9$      &$-0.042$ &$-0.036$ &$-0.011$ &$-0.003$ \\
\hline
$2$      &$-0.029$ &$-0.027$ &$0.217$  &$0.152$ \\
$4$      &$-0.020$ &$-0.018$ &$0.093$  &$0.051$ \\
$6$      &$-0.014$ &$-0.012$ &$0.049$  &$0.023$ \\
$8$      &$-0.010$ &$-0.009$ &$0.030$  &$0.012$ \\
$10$     &$-0.008$ &$-0.006$ &$0.019$  &$0.007$ \\
\hline
\end{tabular}
\end{center}
\end{table}
The numerical results show that for both odd and even-order $\xi$-moments, their magnitudes exhibit a significant decreasing trend as the $n$-order increases. This asymptotic behavior is an expected characteristic of normalized and bounded LCDAs, indicating that the contribution of higher order moments to LCDAs will gradually reduce. This behaviours is consistent with the convergence conditions required for physical LCDAs. The QCDSR'05 provides the $\xi$-moments $\langle\xi^1\rangle_{2;K_0^*(700)}=-0.55$ and $\langle\xi^3\rangle_{2;K_0^*(700)}=-0.21$. Our results are in good agreement with these values. In order to be consistent with the matrix element definition of twist-3 LCDAs, we have eliminated the decay constant $\bar{f}_{K_0^*(700)}$ in Ref.~\cite{Cheng:2005nb}, $i.e.,$ $\phi_{3;K_0^*(700)}^{p}(x, \mu_0)=1$ and $\phi_{3;K_0^*(700)}^{\sigma}(x, \mu_0)=6x\bar{x}$. Using Eq.~\eqref{eq:xi} to calculated the first four moments, it can be observed that the odd-order moments are zero and even-order moments satisfy $\langle\xi^{2,4}\rangle_{3;K_0^*(700)}^{p}\approx 0.333(0.200)$ and $\langle\xi^{2,4}\rangle_{3;K_0^*(700)}^{\sigma}\approx 0.200(0.086)$. However, within the framework of TF model, the odd-order moments will result in smaller non-zero values, while the corresponding even-order moments are also slightly shifted.

\subsection{The TFFs for the $\bar{B}_s^0\to K_0^*(700)$ transition}
To calculate the TFFs for the $\bar{B}_s^0\to K_0^*(700)$ transition, we adopt the following parameters: the $K_0^*(700)$ state mass $m_{K_0^*(700)}=838\pm11 \,\rm{MeV}$, the $\bar{B}_s^0$-meson mass $m_{\bar{B}_s^0}=5366.93\pm0.10 \,\rm{MeV}$, the $s$-quark current mass $m_s = 93_{-5}^{+11} \, \rm{MeV}$ at $\mu = 2 \,\rm{GeV}$, and the $b$-quark pole mass $m_b(\mu_b)=4.18^{+0.03}_{-0.02}\,\rm{GeV}$~\cite{ParticleDataGroup:2024cfk}. The $B_s$-meson decay constant is taken as $f_{B_s}=0.266\pm0.019 \, \rm{GeV}$, while the decay constant of the $K_0^*(700)$ state is set to $\bar{f}_{K_0^*(700)} = 340\pm20 \, \rm{MeV}$ at the initial scale $\mu_0= 1 \, {\rm GeV}$. Both values are taken from QCDSR calculations~\cite{Cheng:2005nb}. The effective threshold parameter is chosen as $s_{B_s}=36\pm1 \, {\rm GeV}$. The Borel windows are chosen as $13 \, {\rm GeV^2}\leq M^2\leq 15 \, {\rm GeV^2}$ for the $f_+^{\bar B_s^0 K_0^*(700)}(q^2)$, and $29 \, {\rm GeV^2}\leq M^2\leq 31 \, {\rm GeV^2}$ for $f_{-, {\rm T}}^{\bar B_s^0 K_0^*(700)}(q^2)$. For the $\bar{B}_s^0 \to K_0^*(700)$ transition, the energy scale $\mu$ is set to the typical momentum transfer of the process, to properly separate long-distance and short-distance contributions. The constituent quarks of scalar states are inherently off-shell, far from their mass shell, and thus carry a non-zero virtuality whose magnitude is of the order $m_{S}^2$, where $m_S$ denotes the scalar states mass. These constituent quarks carry the total momentum of the scalar states, which corresponds to a large longitudinal light-cone momentum. In order to consistently describe the hadronic structure with large intrinsic virtuality, the LCDAs must be evaluated at a normalization scale satisfying $\mu\geq m_S$. In this regime, the off-shell degrees of freedom of the constituent quarks can be included in perturbative calculations as effective dynamical variables. Conversely, when $\mu < m_S$ is chosen, these off-shell modes are partially or fully integrated out, preventing a well defined perturbative QCD expansion. Accordingly, we adopt the scale $\mu=\sqrt{m_{\bar{B}_s^0}^2-m_b^2}\simeq 3.0 \, {\rm GeV}$ for this work. Thus, the LCDAs parameters must be evolved to the corresponding process energy scale. This scale evolution is performed by solving the renormalization group equations~\cite{Yang:1993bp, Hwang:1994vp}, ensuring the consistency and accuracy of physical quantities across different scales. Using the above parameters, Eqs.~\eqref{eq:LCSRTFFs1}-\eqref{eq:LCSRTFFs3}, and including all sources of uncertainty, the $q^2$-dependence of $\bar{B}_s^0 \to K_0^*(700)$ TFFs in the allowable LCSR kinematic region is obtained. In particular, the values of those TFFs at the large recoil point $q^2=0$ are listed in Table~\ref{table:TFFsBskappa}.
\begin{table}[htb]
\footnotesize
\begin{center}
\renewcommand{\arraystretch}{1.2}
\setlength{\tabcolsep}{23pt}
\caption{Numerical results for the $\bar{B}_s^0 \to K_0^*(700)$ TFFs at the large recoil point $i.e.,$ $f_{\pm,{\rm T}}^{\bar B_s^0 K_0^*(700)}(0)$ for two schemes. Predictions from PQCD'08~\cite{Li:2008tk}, LCSR~\cite{Sun:2010nv,Wang:2014vra} and CQM'15~\cite{Issadykov:2015iba} are also listed for comparison.}
\label{table:TFFsBskappa}
\begin{tabular}{l l l l}
\hline
Method  &$f_+^{\bar B_s^0 K_0^*(700)}(0)$ &$f_-^{\bar B_s^0 K_0^*(700)}(0)$  &$f_{\rm T}^{\bar B_s^0 K_0^*(700)}(0)$ \\
\hline
This Work(S1)                          &$0.431^{+0.045}_{-0.043}$  &$-0.397^{+0.038}_{-0.039}$  &$0.488^{+0.048}_{-0.047}$  \\
This Work(S2)                          &$0.432^{+0.045}_{-0.043}$  &$-0.398^{+0.038}_{-0.040}$  &$0.489^{+0.049}_{-0.047}$  \\
PQCD'08 \cite{Li:2008tk}               &$0.29^{+0.07}_{-0.06}$     &$-0.29^{+0.07}_{-0.06}$     &$0.31^{+0.07}_{-0.06}$  \\
LCSR'10 \cite{Sun:2010nv}              &$0.53$                     &$-0.53$                     &$-$\\
LCSR'14 \cite{Wang:2014vra}            &$0.442^{+0.033}_{-0.033}$  &$-0.340^{+0.030}_{-0.030}$  &$0.596^{+0.049}_{-0.049}$  \\
LCSR'14(LO) \cite{Wang:2014vra}        &$0.36$                     &$-0.29$                     &$0.42$  \\
CQM'15(I)  \cite{Issadykov:2015iba}    &$0.138$                    &$-0.138$                    &$-$  \\
CQM'15(II)  \cite{Issadykov:2015iba}   &$0.274$                    &$-0.268$                    &$-$  \\
\hline
\end{tabular}
\end{center}
\end{table}
Specifically, we calculate the TFFs for the $\bar{B}_s^0 \to K_0^*(700)$ transition using the two LCDA schemes (S1) and (S2), corresponding to $\phi_{2;K_0^*(700)}^{\rm(S1)}(x,\mu)$ and $\phi_{2;K_0^*(700)}^{\rm(S2)}(x,\mu)$, respectively. For comparison, we also presents the calculation results of other theoretical approaches, including PQCD'08 \cite{Li:2008tk}, LCSR~\cite{Sun:2010nv, Wang:2014upa, Wang:2014vra} and CQM~\cite{Issadykov:2015iba}. The CQM yields two sets of results corresponding to $\Lambda_S=0.8$ and $\Lambda_S=1.5$, which we label as CQM(I) and CMQ(II), respectively. It is found that the large recoil values of TFFs obtained from the two schemes are nearly identical, as this result is governed by the full range convolution of the LCDAs with the exponential kernel. Although schemes (S1) and (S2) exhibit minor discrepancies in the peak and local regions, the exponential kernel suppresses these differences, leading to mutually canceling integral contributions and consequently nearly identical endpoint values and uncertainties for the TFFs. The TFFs obtained from the two schemes are in excellent agreement, further confirming the stability and reliability of our predictions. A comparative analysis shows that our predictions for $f_{\pm,{\rm T}}^{\bar B_s^0 K_0^*(700)}(0)$ are compatible with those from PQCD and LCSR within their theoretical uncertainties. The differences in predictions from various theoretical frameworks arise from the distinct choices of non-perturbative parameters, renormalization schemes, and phenomenological inputs.

In the LCSR framework, the OPE is valid only for small and intermediate momentum transfer squared $m_\ell^2\leq q^2 < (m_b-m_{K_0^*(700)})^2-2(m_b-m_{K_0^*(700)})\chi$, where extrapolations to large $q^2$ become unreliable. Here, $\chi$ denotes a typical hadronic scale of approximately $0.5 \, \rm{GeV}$. For the $K_0^*(700)$ state, we thus adopt the valid LCSR region $m_\ell^2\leq q^2 \leq 8.4 \, \rm{GeV^2}$. In order to further calculate the differential decay widths and branching fractions of four-body semileptonic $\bar{B}_s^0 \to K_0^*(700)(\to K^0\pi^+)\ell^-\bar{\nu}_{\ell}$ decay, we must extrapolate the LCSR results to the whole kinematically allowed region $q^2 \in [m_\ell^2, (m_{\bar{B}_s^0}-m_{K_0^*(700)})^2]$. In the paper, we employ the simplified series expansion (SSE)~\cite{Bourrely:2008za}, defined as
\begin{align}
F_i(q^2)=\frac{1}{1-q^2/m^2_{\bar{B}_s^0}}\sum_{k=0, 1, 2}\beta_{k, i}z^k(q^2, t_0),
\end{align}
where $F_i(q^2)$ with $i=(1, 2, 3)$ denotes the TFFs $f_{\pm}^{\bar B_s^0 K_0^*(700)}(q^2)$ and $f_{\rm T}^{\bar B_s^0 K_0^*(700)}(q^2)$, respectively. The function
\begin{align}
z^k(q^2, t_0)=\frac{\sqrt{t_+-q^2}-\sqrt{t_+-t_0}}{\sqrt{t_+-q^2}+\sqrt{t_+-t_0}}
\end{align}
is the standard conformal mapping variable used in the $z$-expansion framework for parameterizing TFFs, where $t_{\pm}=(m_{\bar{B}_s^0} \pm m_{K_0^*(700)})^2$ and $t_0=t_+(1-\sqrt{1-t_-/t_+})$. The real expansion coefficients $\beta_{k, i}$ are determined by requiring the extrapolation quality to satisfy $\Delta<1\%$, where $\Delta$ is defined as
\begin{align}
\Delta=\frac{\sum_t|F_i(t)-F^{{\rm fit}}_{i}(t)|}{\sum_t|F_i(t)|}\times 100,
\end{align}
\begin{table}[htb]
\footnotesize
\begin{center}
\renewcommand{\arraystretch}{1.3}
\setlength{\tabcolsep}{17pt}
\caption{The extrapolation parameters $\beta_{k,i}(k=0,1,2)$ and quality of extrapolation $\Delta$ for the semileptonic $\bar{B}_s^0 \to K_0^*(700)$ decays. The obtained results include the numerical values of the upper limit, central value, and lower limit.} \label{table:TFFsExt}
\begin{tabular}{l l l l l}
\hline
&  &$f_+^{\bar{B}_s^0 K_0^*(700)}(q^2)$  &$f_{-}^{\bar{B}_s^0 K_0^*(700)}(q^2)$  &$f_{\rm T}^{\bar{B}_s^0 K_0^*(700)}(q^2)$   \\
\hline
                                    &$\beta_{0,i}$   &$0.476$     &$-0.359$     &$0.536$    \\
                                    &$\beta_{1,i}$   &$-1.227$    &$-0.260$     &$-0.028$   \\
\raisebox{2.5ex}[0pt]{upper limit}  &$\beta_{2,i}$   &$-6.320$    &$1.962$      &$-4.250$   \\
                                    &$\Delta$        &$0.024\%$   &$0.730\%$    &$0.120\%$  \\
\hline
                                    &$\beta_{0,i}$   &$0.431$     &$-0.397$     &$0.488$    \\
                                    &$\beta_{1,i}$   &$-1.164$    &$-0.368$     &$-0.047$   \\
\raisebox{2.5ex}[0pt]{central value}&$\beta_{2,i}$   &$-6.142$    &$1.644$      &$-3.799$   \\
                                    &$\Delta$~~      &$0.026\%$   &$0.680\%$    &$0.180\%$  \\
\hline
                                    &$\beta_{0,i}$   &$0.388$     &$-0.436$     &$0.441$    \\
                                    &$\beta_{1,i}$   &$-1.132$    &$-0.446$     &$-0.113$   \\
\raisebox{2.5ex}[0pt]{lower limit}  &$\beta_{2,i}$   &$-5.958$    &$1.543$      &$-3.656$   \\
                                    &$\Delta$        &$0.031\%$   &$0.632\%$    &$0.242\%$  \\
\hline
\end{tabular}
\end{center}
\end{table}
where $t\in[0, \frac{1}{2}, \cdots, \frac{27}{2}, 14] \, {\rm GeV^2}$. Given the nearly identical results from the two schemes, we present only the predictions from scheme (S1) in the following discussion. The resulting numerical values (including the central values, the upper and lower limits) are listed in Table~\ref{table:TFFsExt}. It can be observed that all fitted parameters satisfy the constraint $\Delta<1\%$ for the extrapolation quality.
\begin{figure}[htb]
\begin{center}
~\includegraphics[width=0.49\textwidth]{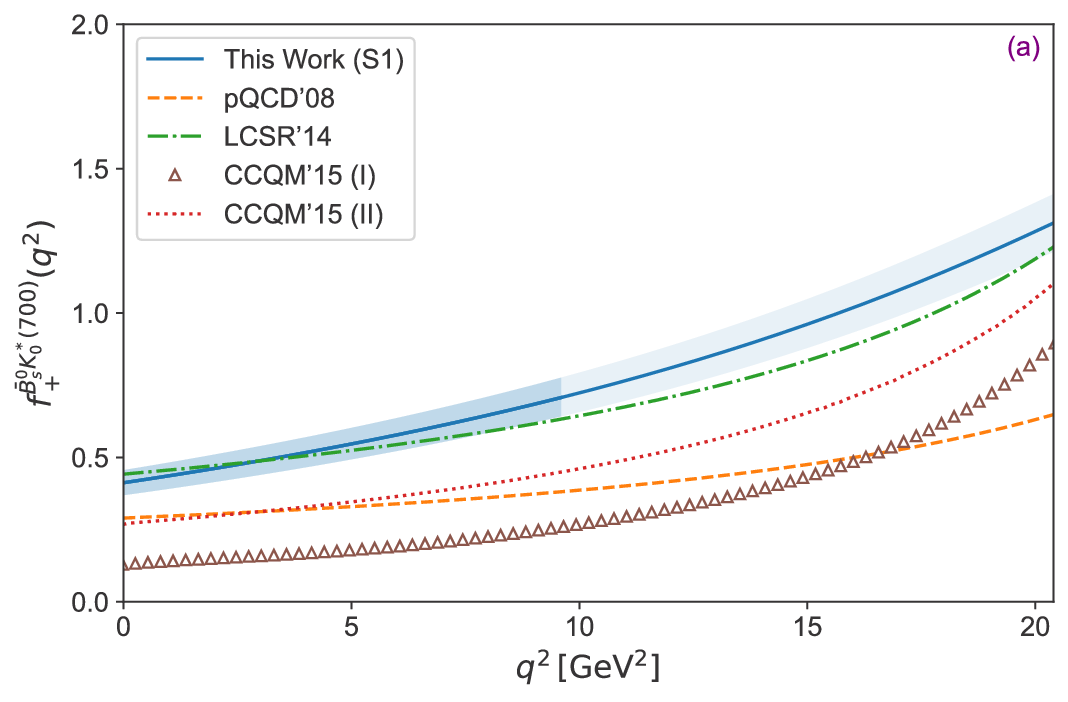}
~\includegraphics[width=0.49\textwidth]{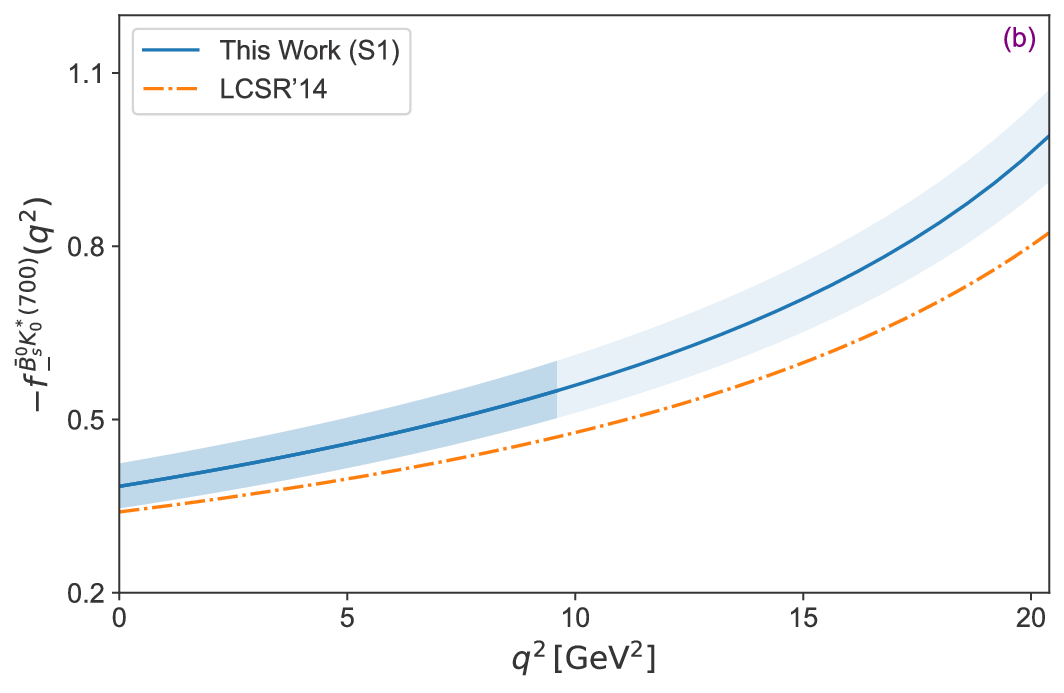}
~\includegraphics[width=0.49\textwidth]{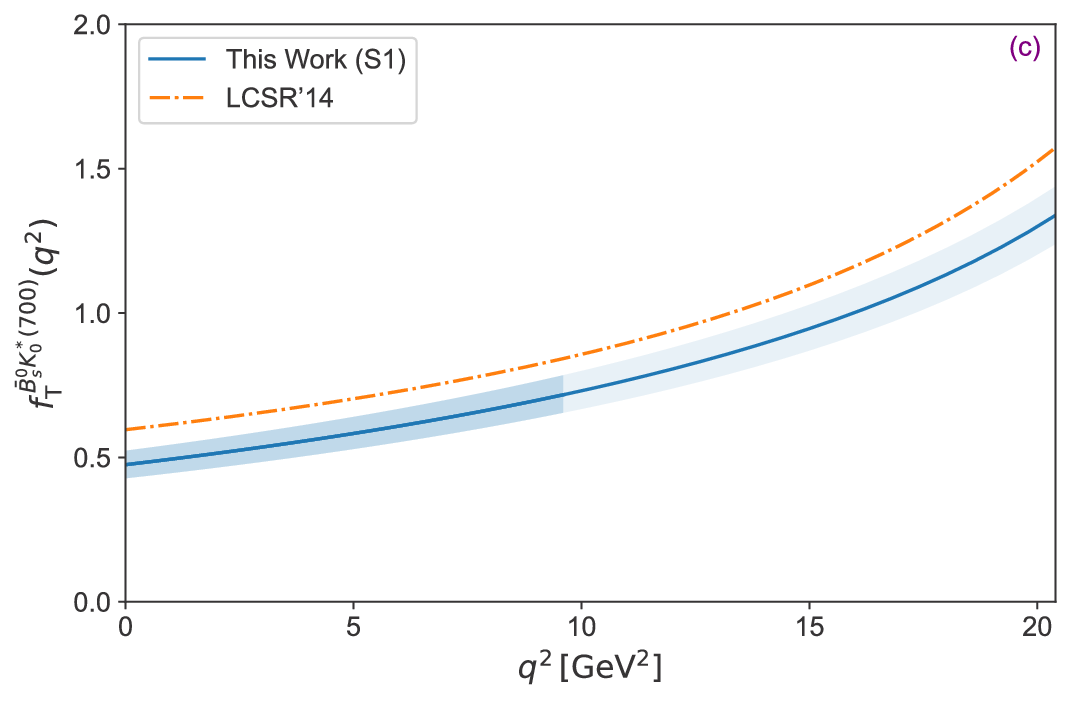}\\
\end{center}
\caption{The behaviors of the TFFs $f_+^{\bar B_s^0 K_0^*(700)}(q^2)$\,(a), $-f_-^{\bar B_s^0 K_0^*(700)}(q^2)$\,(b) and $f_{\rm T}^{\bar B_s^0 K_0^*(700)}(q^2)$\,(c) in the full $q^2$ region. The solid and dashed lines represent the predicted central values, while the shaded bands indicate the corresponding uncertainties. The dark and light parts within the shadow band correspond respectively to the direct LCSR calculation and the extrapolation results. The theoretical predictions of PQCD'08~\cite{Li:2008tk}, LCSR'14~\cite{Wang:2014vra} and CQM'15~\cite{Issadykov:2015iba} were presented for comparison.}
\label{fig:TFFs}
\end{figure}
Then, we present the whole $q^2$-dependence of TFFs $f_\pm^{\bar B_s^0 K_0^*(700)}(q^2)$ and $f_{\rm T}^{\bar B_s^0 K_0^*(700)}(q^2)$ in Fig.~\ref{fig:TFFs}. Here, the solid line shows our central values, and the shaded band corresponds to the uncertainties from the input parameters. The darker shaded regions show the LCSR results, while the lighter bands show the extrapolated results from the SSE method. In particular, the predictions from PQCD'08~\cite{Li:2008tk}, LCSR'14~\cite{Wang:2014vra}, and CQM'15~\cite{Issadykov:2015iba} are also included for comparison. From Fig.~\ref{fig:TFFs}, one can find that, out predictions for $f_{\pm, {\rm T}}^{\bar B_s^0 K_0^*(700)}(q^2)$ exhibits smooth and stable behavior in low-to-intermediate region, and exist slight differences with the LCSR'14 calculations. In contrast, the LCSR theoretical prediction for TFF $f_+^{\bar B_s^0 K_0^*(700)}(q^2)$ shows a significant deviation from the results of PQCD'08 and CQM'15 calculations

\subsection{Decay widths and branching fractions for the semileptonic $\bar{B}_s^0\to K_0^*(700)(\to K^0\pi^+)\ell^-\bar{\nu}_\ell$ decay}

Combining the input parameters with Eq.~\eqref{eq:DDW}, we can calculate the differential decay widths of four-body semileptonic $\bar{B}_s^0 \to K_0^*(700)(\to K^0\pi^+)\ell^-\bar{\nu}_{\ell}$ decay. In the calculation, we take the Fermi coupling constant $G_F=1.166\times 10^{-5}\,{\rm GeV^{-2}}$, CKM matrix element $|V_{ub}|=(3.82\pm0.20)\times10^{-3}$, and the masses $m_e=0.511\,{\rm MeV}$, $m_\mu=105.658\,{\rm MeV}$, $m_\tau=1776.93\,{\rm MeV}$, $m_{K^0}=497.611\,{\rm MeV}$, and $m_{\pi^+}=139.570\,{\rm MeV}$~\cite{ParticleDataGroup:2024cfk}. The strong coupling constant $g_{K_0^*(700) K^0\pi^+}$ are determined from the measured partial width through the relation~\cite{Chua:2008zza}, $i.e.,$ $\Gamma_{K_0^*(700)}=\frac{p_c}{8\pi m_{K_0^*(700)}^2}g_{K_0^*(700) K\pi}^2$, where $p_c$ is the momentum of $K$ or $\pi$ in the rest frame of $K_0^*(700)$. We extract the strong coupling constant as $g_{K_0^*(700) K^0\pi^+} \approx 5.79\,{\rm GeV}$.

\begin{figure}[htb]
\begin{center}
~\includegraphics[width=0.49\textwidth]{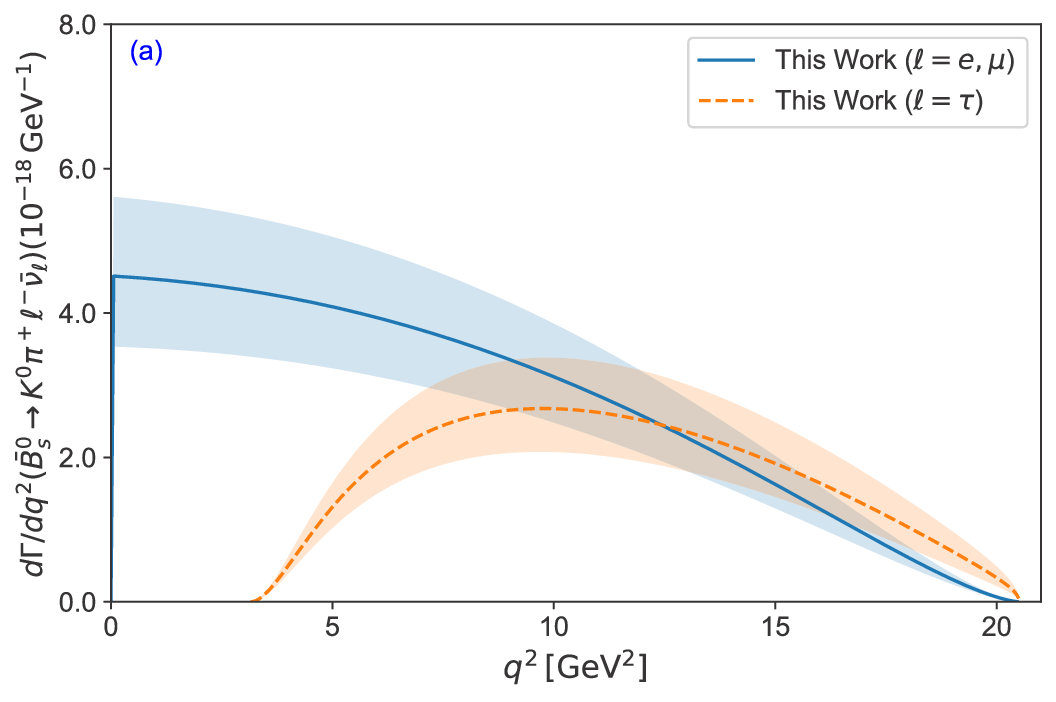}
~\includegraphics[width=0.49\textwidth]{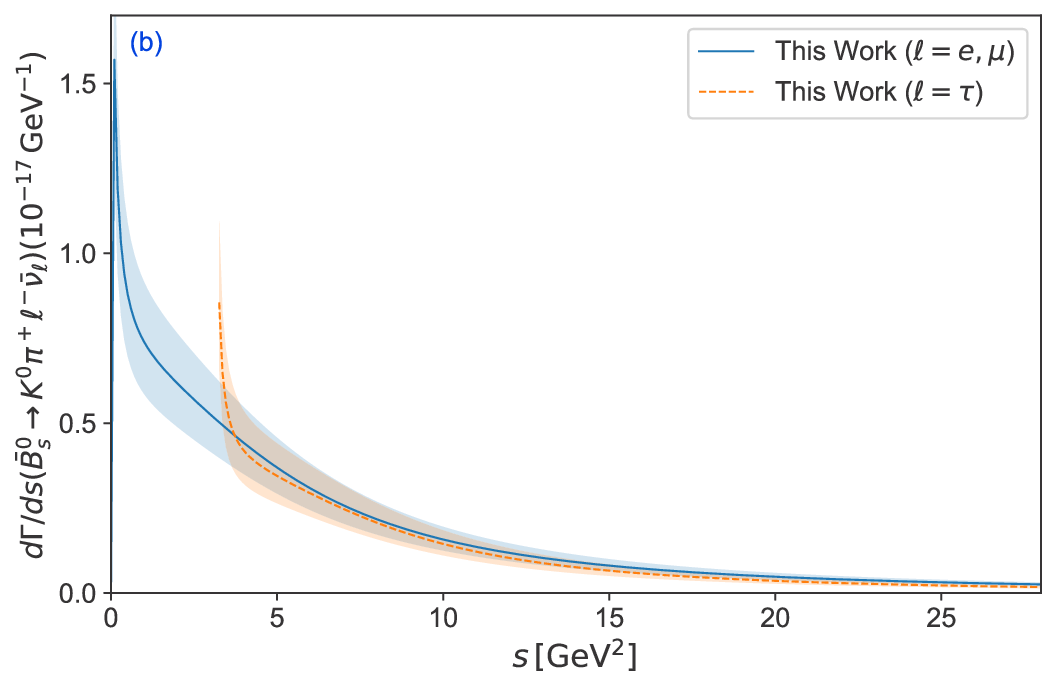}
\end{center}
\caption{The differential decay widths of the four-body semileptonic process $\bar{B}_s^0 \to K_0^*(700)(\to K^0\pi^+)\ell^-\bar{\nu}_\ell$ with $\ell=(e,\mu,\tau)$ are shown as functions of $q^2$ and $s$, illustrating their respective dependence on these kinematic variables. The solid and dashed curves indicate the central values, with the shaded bands representing the corresponding uncertainties.}
\label{fig:DDWs}
\end{figure}

The behaviors of differential decay widths for the four-body semileptonic $\bar{B}_s^0 \to K_0^*(700)(\to K^0\pi^+)\ell^-\bar{\nu}_\ell$ decay versus $q^2$ and $s$ are shown in Fig.~\ref{fig:DDWs}. Here, the solid and dashed curves indicate the central values, with the shaded bands representing the corresponding uncertainties.
In Fig.~\ref{fig:DDWs}\textcolor{blue}{(a)}, for electron (or muon)-channel, the phase space suppression effect is relatively weak in the small $q^2$ region, resulting in reversely high decay ratios. As $q^2$ increases, the phase space compression effect gradually becomes dominant, effectively suppressing the contributions from the TFFs. This leads to a monotonic decrease in the decay widths, which approaches zero at the endpoint $q^2=(m_{\bar{B}_s^0}-m_{K_0^*(700)})^2$. For $\tau$-channel, the large mass reduces the available kinetic energy in the hadron final state, significantly compressing the phase space and thereby causing a significant decrease in the decay width. The peak value of the differential decay width $d\Gamma/ds$ occurs at $s=(m_K+m_\pi)^2$. From Fig.~\ref{fig:DDWs}\textcolor{blue}{(b)}, it can be observed that the width becomes less dependent on $s$ for $s \gtrsim 15\,{\rm GeV}^2$. This behavior indicates that when $s$ approaches the maximum invariant mass allowed by the hadron system, the phase space factor and resonance spectral function $P(s)$ will simultaneously be depressed.

Furthermore, by using the extrapolated TFFs within the SSE approach framework and combining with the $\bar{B}_s^0$ meson decay lifetime, we calculated the four-body semileptonic decays $\mathcal{B}(\bar{B}_s^0 \to K^0\pi^+ \ell^-\bar{\nu}_\ell)$ for the $S$-wave $K_0^*(700)$ resonance and the three-body semileptonic $\mathcal{B}(\bar{B}_s^0 \to K_0^*(700)\ell^-\bar{\nu}_\ell)$ decays under the narrow width approximations is also calculated. There results are listed in Table~\ref{table:BRs}.
\begin{table}[htb]
\footnotesize
\begin{center}
\renewcommand{\arraystretch}{1.5}
\setlength{\tabcolsep}{10pt}
\caption{The branching fractions $(10^{-4})$ of the four-body $\bar{B}_s^0 \to K_0^*(700)(\to K^0\pi^+)\ell^-\bar{\nu}_\ell$ and the three-body $\bar{B}_s^0 \to K_0^*(700)\ell^-\bar{\nu}_\ell$ semileptonic decays with $\ell=(e,\mu,\tau)$, respectively. For comparison, we also present the theoretical predictions of LCSR'14~\cite{Wang:2014upa}, CQM'15(II)~\cite{Issadykov:2015iba}, and PQCD'08~\cite{Li:2008tk}.}
\label{table:BRs}
\begin{tabular}{l c c c c}
\hline
Decay Modes  &This Work  &LCSR'14~\cite{Wang:2014upa} &CQM'15(II)~\cite{Issadykov:2015iba} &PQCD'08~\cite{Li:2008tk}\\
\hline
$\bar{B}_s^0 \to K_0^*(700)(\to K^0\pi^+)e^-\bar{\nu}_e$         &$1.290^{+0.450}_{-0.345}$  &$1.02\pm0.15$  &$0.61$     &$0.70^{+0.40}_{-0.26}$\\
$\bar{B}_s^0 \to K_0^*(700)(\to K^0\pi^+)\mu^-\bar{\nu}_\mu$     &$1.288^{+0.449}_{-0.345}$  &$1.02\pm0.15$  &$0.61$     &$0.70^{+0.40}_{-0.26}$\\
$\bar{B}_s^0 \to K_0^*(700)(\to K^0\pi^+)\tau^-\bar{\nu}_\tau$   &$0.703^{+0.308}_{-0.223}$  &$0.53\pm0.09$  &$0.12$     &$0.43^{+0.25}_{-0.16}$\\
$\bar{B}_s^0 \to K_0^*(700)e^-\bar{\nu}_e$                       &$2.544^{+0.887}_{-0.680}$  &$2.06\pm0.31$  &$1.23$  &$1.42^{+0.82}_{-0.53}$\\
$\bar{B}_s^0 \to K_0^*(700)\mu^-\bar{\nu}_\mu$                   &$2.539^{+0.886}_{-0.680}$  &$2.06\pm0.31$  &$1.23$  &$1.42^{+0.82}_{-0.53}$\\
$\bar{B}_s^0 \to K_0^*(700)\tau^-\bar{\nu}_\tau$                 &$1.386^{+0.607}_{-0.440}$  &$1.07\pm0.19$  &$0.25$  &$0.88^{+0.52}_{-0.33}$\\
\hline
\end{tabular}
\end{center}
\end{table}
We also present other predictions from LCSR'14~\cite{Wang:2014upa}, CQM'15(II)~\cite{Issadykov:2015iba}, and PQCD'08~\cite{Li:2008tk} for comparison. In the calculation, we take into account all the uncertainties caused by the input parameters. As far as we know, the four-body semileptonic $\bar{B}_s^0 \to K_0^*(700)(\to K^0\pi^+)\ell^-\bar{\nu}_\ell$ decays has not been studied before. Therefore, we conducted further theoretical analysis by utilizing the branching fractions of the three-body semileptonic $\bar{B}_s^0 \to K_0^*(700)\ell^-\bar{\nu}_\ell$ decays and the basic input parameters reported in LCSR'14~\cite{Wang:2014upa}, CQM'15(II)~\cite{Issadykov:2015iba} and PQCD'08~\cite{Li:2008tk}. We perform the corresponding theoretical calculations by adopting the relativistic Flatt\'e formula, so as to carry out comparisons and analyses among different decay modes. The research results show that the four-body decay branching fractions obtained by using the Flatt\'e formula to describe the $K_0^*(700) \to K^0\pi^+$ resonance decay is significantly smaller than the results under the narrow width approximation. Compared with the three-body decays under narrow width approximation, the four-body decays results in a greater number of final state particles, leading to a significant reduction in the available phase space. Meanwhile, the four-body semileptonic decay studied in this work is a cascading precess via an $K_0^*(700)$ resonance, and there is a decay width of $K_0^*(700) \to K^0\pi^+$ involved. Furthermore, the Flatt\'e formula takes into account the final state strong interaction effect and resonant linear correction of the $K^0\pi^+$ system, making the calculation results consistent with the actual physical process. Our results are in excellent agreement within the uncertainty range with the predictions of LCSR'14~\cite{Wang:2014upa}, while the differences from PQCD'08~\cite{Li:2008tk} and CQM'15(II)~\cite{Issadykov:2015iba} mainly stem from TFFs. Due to lepton universality, the branching fractions of the electron and muon channels are almost identical. The minor differences arise from the slight phase space effects caused by the mass difference between electrons and muons.

\section{SUMMARY}\label{IV}
Because the strange $K_0^*(700)/\kappa$ state has a controversial internal structure, a relatively large width ($\Gamma_{K_0^*(700)}=463\pm27\,\rm{MeV}$), and significant non-perturbative characteristics, it cannot be treated as a conventional narrow resonance. Therefore, we treat the $K_0^*(700)$ state as an intermediate resonance, and describe its invariant mass distribution using the relativistic Flatt\'e formula, thereby studying the four-body semileptonic decays $\bar{B}_s^0 \to K^0\pi^+\ell^-\bar{\nu}_\ell$ with $\ell=(e,\mu,\tau)$ for the $S$-wave $K_0^*(700)$ resonance. In this paper, the $K_0^*(700)$ resonance is described as a $q\bar{q}$ state, and we employ the BHL prescription to construct an LCHO model for LCDAs $\phi_{2;K_0^*(700)}(x, \mu)$ and $\phi_{3;K_0^*(700)}^{p;\sigma}(x, \mu)$. Based on the constructed LCHO model and taking into account the relevant constraints, we proposed the two scheme for the twist-2 LCDAs, {\it i.e.} $\phi_{2;K_0^*(700)}^{\rm (S1)}(x, \mu)$ and $\phi_{2;K_0^*(700)}^{\rm (S2)}(x, \mu)$. The model parameters are given in Eqs.~\eqref{eq:T2}-\eqref{eq:T3}. In Fig.~\ref{fig:phi}, we present the behaviors of twist-2 and twist-3 LCDAs. In schemes (S1) and (S2), the twist-2 LCDA exhibit a single peak behaviors and antisymmetry, which is consistent with the result of QCDSR. Furthermore, we have calculated the first ten-order $\xi$-moments at the scale $\mu_0=1\,{\rm GeV}$, which are presented in Table~\ref{table:xin}. The results show that as the $n$-th increases, the magnitudes of both odd and even $\xi$-moments exhibit a decreasing trend, and the contributions of higher order $\xi$-moments become increasingly smaller.

Subsequently, we calculate the TFFs $f_\pm^{\bar B_s^0 K_0^*(700)}(0)$ and $f_{\rm T}^{\bar B_s^0 K_0^*(700)}(0)$, and the numerical results are shown in Table~\ref{table:TFFsBskappa}. It is observed that our predicted results for $f_{\pm, {\rm T}}^{\bar B_s^0 K_0^*(700)}(0)$ are close to the PQCD and LCSR results in the large recoil region. Due to the results of the two schemes are highly consistent, we choose scheme (S1) for the subsequent discussions. Considering that the LCSR approach is only applicable in the small and intermediate region, we employ the SSE parametrization to investigate the behaviours of TFFs in the full $q^2$ region. The extrapolation parameters $\beta_{k, i}(k=0, 1, 2)$ and quality of extrapolation $\Delta$ for the $\bar{B}_s^0 \to K_0^*(700)$ transition, as shown in Table~\ref{table:TFFsExt}. The behaviours of the TFFs in the full $q^2$ region are depicted in Fig.~\ref{fig:TFFs}. These the TFFs $f_{\pm, {\rm T}}^{\bar B_s^0 K_0^*(700)}(q^2)$ show smooth and stable behavior in the small and intermediate region, and exist slight differences with the LCSR'14 calculations.

Finally, we calculated the differential decay widths and branching fractions for the four-body semileptonic decays $\bar{B}_s^0 \to K_0^*(700)(\to K^0\pi^+)\ell^-\bar{\nu}_\ell$ and the three-body semileptonic decays $\bar{B}_s^0 \to K_0^*(700)\ell^-\bar{\nu}_\ell$ under the narrow width approximation, with $\ell=(e,\mu,\tau)$, as shown in Fig.~\ref{fig:DDWs} and Table~\ref{table:BRs}. Our results of the branching fractions are in good agreement with those from LCSR'14 within uncertainties. This paper has expanded the research on the four-body semileptonic decays precess involving the strange $K_0^*(700)$ state as an intermediate resonance. We hope that the predicted results obtained will provide unique insights for future experimental and theoretical research.

\section{ACKNOWLEDGMENTS}\label{V}
We are grateful to Prof. Tao Zhong for helpful discussions. This work was supported in part by the National Natural Science Foundation of China under Grant No.12265010, the Project of Guizhou Provincial Department of Science and Technology under Grants No.MS[2025]219 and No.CXTD[2025]030.

\end{document}